\documentclass[ aps, twocolumn, nofootinbib, showkeys, notitlepage, superscriptaddress]{revtex4-2}

\usepackage{natbib}
\usepackage{lipsum}
\usepackage{graphicx}
\usepackage{dcolumn}
\usepackage{bm}
\usepackage{amsmath}
\usepackage{float}
\usepackage{multirow}
\usepackage{slashed}
\usepackage{braket}
\usepackage{xcolor}
\usepackage{physics}
\usepackage{multirow}
\usepackage{gensymb}
\usepackage[colorlinks=true, pdfstartview=FitV, bookmarks=true, bookmarksnumbered=true, breaklinks]{hyperref}
\usepackage{mathtools,braket}
\usepackage{lipsum}  
\usepackage{color}
\usepackage[normalem]{ulem}   
\renewcommand{\sout}{\bgroup \color{red} \ULdepth=-.5ex \ULset}
\definecolor{blue}{rgb}{0.0, 0.0, 1.0}
\definecolor{red}{rgb}{1.0, 0.0, 0.0}
\definecolor{royalblue}{rgb}{0.0, 0.14, 0.4}
\hypersetup{linkcolor=blue, citecolor=blue, urlcolor=blue}

\def\orcid#1{\kern .08em\href{https://orcid.org/#1}{\includegraphics[keepaspectratio,width=0.7em]{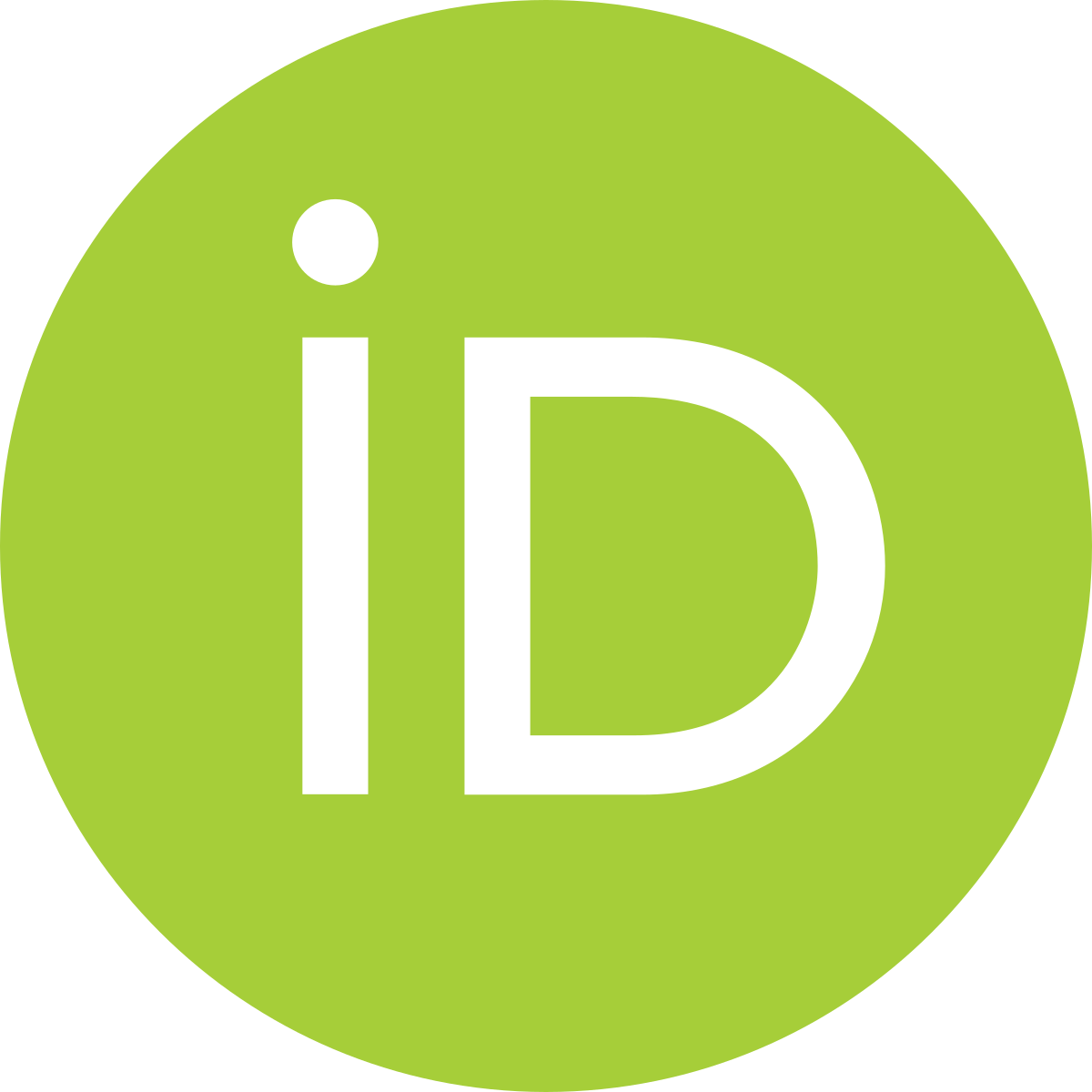}}}

\begin{document}

\title{Polarization-dependent mass modifications of $\phi$ meson with finite momentum\\ in nuclear matter}

\author{Ahmad Jafar Arifi\orcid{0000-0002-9530-8993}}
\email{aj.arifi01@gmail.com}
\affiliation{Advanced Science Research Center, Japan Atomic Energy Agency, Tokai, Ibaraki 319-1195, Japan}
\affiliation{Research Center for Nuclear Physics, The University of Osaka, Ibaraki, Osaka 567-0047, Japan}

\author{Philipp Gubler\orcid{0000-0002-0991-8462}}
\email{gubler.philipp@jaea.go.jp}
\email{philipp.gubler1@gmail.com}
\affiliation{Advanced Science Research Center, Japan Atomic Energy Agency, Tokai, Ibaraki 319-1195, Japan}

\author{Kazuo Tsushima\orcid{0000-0003-4926-1829}}
\email{kazuo.tsushima@cruzeirodosul.edu.br}
\email{kazuo.tsushima@gmail.com}
\affiliation{Laboratório de Física Teórica e Computacional-LFTC, 
Programa de P\'{o}sgradua\c{c}\~{a}o em Astrof\'{i}sica e F\'{i}sica Computacional, Universidade Cidade de S\~{a}o Paulo, 01506-000 S\~{a}o Paulo, SP, Brazil}

\date{\today}

\begin{abstract}

We investigate the in-medium properties of the $\phi$ meson with finite momentum, going beyond the commonly studied case at rest. 
In a nuclear medium, Lorentz invariance is broken, leading to distinct longitudinal and transverse polarization modes that evolve differently with density and momentum. 
Within an effective Lagrangian approach, we calculate the polarization-dependent mass shifts and width modifications of the $\phi$ meson arising from $K\bar{K}$ loops and mean-field interactions.
The divergent loop integrals are regulated using two different schemes: a covariant form factor and dimensional regularization.
Our results show that the mass shift of the transverse polarization is independent of the $\phi$-meson momentum, whereas that of the longitudinal polarization decreases quadratically with momentum. 
This difference originates from the coupling of the longitudinal mode to the vector mean field and derivative-type interactions in the self-energy. 
These effects have direct implications for experimental observables, especially for upcoming measurements at J-PARC, and provide a new prediction for experiments studying hadron dynamics in dense matter.

\end{abstract}

\maketitle

\section{Introduction}

Understanding hadron property changes in dense nuclear environments is a central challenge in nonperturbative quantum chromodynamics (QCD), as medium-induced modifications reflect changes in quark and gluon dynamics and the reduction of the quark condensate, signaling partial restoration of chiral symmetry (in light quark sector)~\cite{Brown:1995qt, Leupold:2009kz, Hayano:2008vn}. Light vector mesons are especially important, since their masses and spectral functions directly probe such effects. Early studies, notably Brown-Rho scaling~\cite{Brown:1991kk} and QCD sum rules (QCDSR)~\cite{Hatsuda:1991ez}, predicted downward mass shifts with increasing density, triggering extensive theoretical and experimental efforts. Among them, the $\phi$ meson is particularly notable: its dominant $s\bar{s}$ structure and narrow vacuum width make it a relatively clean probe of in-medium dynamics, less affected by conventional hadronic many-body effects than the $\rho$ and $\omega$ mesons.

The in-medium properties of the $\phi$ meson are particularly sensitive to the strange sigma term of the nucleon~\cite{Gubler:2014pta}, $\sigma_{sN}\equiv m_s\langle N|\bar{s}s|N\rangle$, which controls the density dependence of the strange quark condensate~\cite{Gubler:2016itj, Kim:2022eku}. 
Lattice QCD has also placed constraints on $\sigma_{sN}$~\cite{Durr:2015dna, Yang:2015uis, Abdel-Rehim:2016won, Borsanyi:2020bpd, RQCD:2022xux, FlavourLatticeAveragingGroupFLAG:2024oxs}, although there is a tension with the extracted value from experiments. Studies of the $\phi N$ interaction~\cite{Oh:2001bq, Kim:2021adl, Lyu:2022imf, Abreu:2024qqo} also reveal coupled-channel dynamics, hidden strangeness, and near-threshold effects, motivating searches for $\phi$–nucleus bound states~\cite{Metag:2017yuh,Tolos:2020aln}. Furthermore, transport simulations~\cite{Cassing:2009vt} have been performed to connect theoretical predictions with observables in proton- and photon-induced nuclear reactions~\cite{Muhlich:2002tu, Gubler:2024ovg, Balassa:2025gwv, KEK-PSE325:2025fms}, while in-medium $\phi$ mesons also impact heavy-ion collisions~\cite{Ko:1991kw, Chung:1997mp, Chung:1998ev, Pal:2002aw, E917:2003gec, STAR:2008bgi, Song:2022jcj, Steinheimer:2025mho}.

Experimentally, measurements at various facilities have provided evidence for density-dependent modifications of the $\phi$ meson spectral function~\cite{Ishikawa:2004id, KEK-PS-E325:2005wbm, E325:2006ioe, CLAS:2010pxs, Polyanskiy:2010tj, Hartmann:2012ia, HADES:2018qkj}. While dilepton decays provide a clean and direct probe of in-medium effects, their small branching ratio limits the available statistics; in contrast, the $K\bar{K}$ decay channel provides higher yields but suffers from strong final-state interactions. In particular, the KEK-E325 experiment observed a downward mass shift and substantial broadening via dilepton measurements~\cite{KEK-PS-E325:2005wbm, KEK-PSE325:2025fms}. 
However, the existing experimental results remain partially inconsistent~\cite{KEK-PSE325:2025fms}, preventing a definitive conclusion.
Ongoing and future experiments at J-PARC, in particular, the E16 and E88/Sa$\phi$re programs, aim to significantly improve precision~\cite{YokkaichiE16Proposal, Naruki:2012eka, Aoki:2023qgl, Aoki:2024ood, Sako:2024oxb}. Additional experimental programs on the $\phi$ meson are also planned at GSI~\cite{Messchendorp:2025men}.

On the theoretical side, most studies have focused on the $\phi$ meson at rest using various approaches, e.g.,~\cite{Ko:1992tp, Asakawa:1994tp, Klingl:1997tm, Zschocke:2002mn, Cabrera:2002hc, Cabrera:2016rnc, Cobos-Martinez:2017vtr, Ahmad:2024ohm, Mondal:2025qxm, Kaur:2025kjk}, generally predicting a downward mass shift accompanied by a significant broadening of the decay width in nuclear matter. However, in realistic experiments~\cite{KEK-PS-E325:2005wbm, KEK-PSE325:2025fms}, $\phi$ mesons carry \emph{finite momentum} in a Lorentz symmetry broken environment, generating distinct longitudinal and transverse polarization modes with different in-medium evolution. Such \emph{polarization-dependent effects} represent a qualitatively new aspect of vector meson dynamics and have been explored mainly within QCDSR analyses~\cite{Lee:1997zta, Leupold:1998bt, Kim:2019ybi}, whereas systematic studies within effective hadronic models remain relatively limited.

In this work, we study the in-medium $\phi$ meson properties at finite momentum, with particular emphasis on the $\phi$-polarization effects. 
To this end, we investigate the in-medium $\phi$ meson self-energy arising from $K\bar{K}$ loops within an effective Lagrangian approach~\cite{Ko:1992tp, Cobos-Martinez:2017vtr, Mondal:2025qxm}, combined with in-medium kaon properties obtained from the quark–meson coupling (QMC) model~\cite{Tsushima:1997df}. 
While the previous study~\cite{Cobos-Martinez:2017vtr} employed three-momentum form factors to evaluate the $K\bar{K}$ loops, here we adopt two Lorentz invariant regularization schemes, \emph{covariant form factors} and \emph{dimensional regularization}, to consistently extend the analysis to finite momentum without introducing artificial Lorentz breaking effects.

Our results reveal a quadratic mass decrease with increasing momentum in the longitudinal mode, driven by vector mean fields and derivative interactions, whereas the transverse mass remains essentially unchanged. The behavior of the longitudinal mass is consistent with QCDSR predictions~\cite{Kim:2019ybi}, whereas the transverse mass exhibits a qualitatively different trend. These differences can be resolved in future experiments and provide clear experimental signatures, potentially accessible through angular correlations in the $\phi\to K\bar{K}$ decay channel~\cite{Park:2022ayr}. The planned J-PARC E88/Sa$\phi$re experiment offers a promising opportunity to directly observe polarization-dependent mass splitting~\cite{Sako:2024oxb}.

This paper is organized as follows.
In Sec.~\ref{sec:self-energy}, we introduce the formalism for the self-energy of the $\phi$ meson, including its vacuum contribution, and describe the in-medium modifications of the kaon mass and energy in the QMC model, which provide essential inputs for the self-energy calculation.
Section~\ref{sec:regularization} presents the evaluation of the $\phi$ meson self-energy using two different regularization schemes.
The model parameters employed in the present study are discussed in Sec.~\ref{sec:parameter}.
The numerical results and their detailed analysis are given in Sec.~\ref{sec:result}.
Finally, our conclusions and outlook are summarized in Sec.~\ref{sec:conclusion}.
Technical details of the form-factor and dimensional regularization methods are provided in Apps.~\ref{app:ff} and~\ref{app:dim-reg}, respectively, where analytic expressions are derived to ensure the correctness of the numerical calculations.

\section{Self-energy of $\Phi$ meson}
\label{sec:self-energy}

In this work, we study in-medium modifications of the $\phi$-meson mass and width in the laboratory frame, as relevant for J-PARC and other experiments~\cite{YokkaichiE16Proposal, Naruki:2012eka, Aoki:2023qgl, Aoki:2024ood, Sako:2024oxb}. 
Thus, we focus on medium effects on the $\phi$-meson self-energy at finite momentum $p^\mu = (p^0, 0, 0, |\bm{p}|)$, where the kaon and antikaon inside the loop are modified. 
In a nuclear medium, Lorentz symmetry is broken due to the existence of a preferred frame defined by the medium four-velocity: $u^\mu = (1, 0, 0, 0)$, leading to a separation between transverse and longitudinal polarization modes. 
The self-energy tensor ${\Pi}^{\mu\nu}$ can then be given as~\cite{Gale:1990pn}
\begin{equation}
    {\Pi}^{\mu\nu}(p^2) = P_T^{\mu\nu} \Pi^T(p^2) + P_L^{\mu\nu} \Pi^L(p^2), 
\end{equation}
where the splitting vanishes in the $\phi$-meson rest frame, in which both modes become degenerate. 
Here, $P_T^{\mu\nu}$ and $P_L^{\mu\nu}$ are the transverse and longitudinal projectors which satisfy
\begin{equation}
    P_T^{\mu\nu} + P_L^{\mu\nu}=  - g^{\mu\nu} + \frac{p^\mu p^\nu}{p^2},
\end{equation}
and can be used to separate tensor structures in in-medium self-energies and propagators according to their polarization modes.
Both projectors satisfy the transversality conditions $P_{L/T}^{\mu\nu} p_{\mu} = 0$ and the following condition ${P_L^\mu}_\mu=\frac{1}{2}{P_T^\mu}_\mu=-1$. 
The projectors have also the property $P_{L/T}^{\mu\sigma}P^{L/T}_{\sigma\nu}=-{P_{\nu,L/T}^\mu}$ and $P_{T}^{\mu\sigma}P^{L}_{\sigma\nu}=0$.
The transverse projector can be expressed as $P_T^{00}=P_T^{0i}= P_T^{i0}=0$ and
\begin{align}
 P_T^{ij} = \delta^{ij} - \frac{\bm{p}^i \bm{p}^j}{\bm{p}^2},
\end{align}
while the longitudinal polarization projector can then be computed as 
\begin{equation}
    P_L^{\mu\nu} =  - g^{\mu\nu} + \frac{p^\mu p^\nu}{p^2} - P_T^{\mu\nu}.
\end{equation} 
The longitudinal and transverse self-energy of the $\phi$ meson can therefore be computed as
\begin{align}
   \Pi_T(p^2) = \frac{1}{2} P^T_{\mu\nu}~ \Pi^{\mu\nu}(p^2),\quad
   \Pi_L(p^2) =  P^L_{\mu\nu} ~ \Pi^{\mu\nu}(p^2).
\end{align}

\subsection{Effective Lagrangian approach}

\begin{figure}[b]
    \centering
    \includegraphics[width=0.3\textwidth]{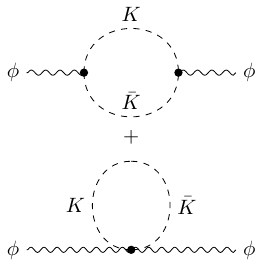} 
    \caption{Self-energy of the $\phi$ meson through: (top) a $K\bar{K}$ loop term and (bottom) a contact four-point coupling term.}
    \label{fig:self-energy}
\end{figure}

The $\phi$-meson self-energy in nuclear matter is evaluated in an effective Lagrangian approach~\cite{Ko:1992tp, Cobos-Martinez:2017vtr}. 
The dominant interaction is provided by the $\phi K\bar K$ coupling,
\begin{equation}\label{eq:lagrangian1}
\mathcal{L}_{\phi K \bar K}
=
i g_\phi \phi^\mu
\left[
\bar K (\partial_\mu K)
-
(\partial_\mu \bar K) K
\right],
\end{equation}
together with the corresponding four-point $\phi\phi K\bar K$ interaction,
\begin{equation}\label{eq:lagrangian2}
\mathcal{L}_{\phi\phi K \bar K}
=
g_\phi^2 \phi^\mu \phi_\mu \bar K K,
\end{equation}
where the kaon and antikaon fields are written as
\begin{equation}
K =
\begin{pmatrix}
K^+ \\
K^0
\end{pmatrix},
\qquad
\bar K =
\begin{pmatrix}
K^- & \bar K^0
\end{pmatrix}.
\end{equation}
The interaction terms in Eqs.~\eqref{eq:lagrangian1} and ~\eqref{eq:lagrangian2} arise naturally from a gauged kaon kinetic Lagrangian, in which the $\phi$ meson is introduced as a vector gauge field~\cite{Ko:1992tp, Cobos-Martinez:2017vtr}. 
Starting from the free kaon Lagrangian $(\partial_\mu K)^\dagger\partial^\mu K$ and promoting the ordinary derivative to a covariant one, $\partial_\mu \to D_\mu = \partial_\mu + i g_\phi \phi_\mu$, the expansion of the gauged kinetic term generates simultaneously the three-point $\phi K\bar K$ vertex and the four-point $\phi\phi K\bar K$ interaction. 
As a consequence, these couplings are not independent but are constrained by local gauge invariance. 

At the level of the $\phi$-meson self-energy, the three-point interaction gives rise to the $K\bar{K}$ loop contribution, while the four-point 
coupling term generates a contact (to be noted by ``con" hereafter) diagram as shown in Fig.~\ref{fig:self-energy}. 
Their combined contribution 
\begin{align}
   \Pi^{\mu\nu}_\mathrm{total}(p^2) = \Pi^{\mu\nu}_{\rm loop}(p^2)+ \Pi^{\mu\nu}_{\rm con}(p^2)
\end{align}
ensures the transversality of the vector-meson self-energy,
\begin{equation}
p_\mu \Pi^{\mu\nu}_\mathrm{total}(p^2) = 0.
\end{equation}

The $\Pi^{\mu\nu}$ computed from a $K\bar{K}$ loop and a contact term in Fig.~\ref{fig:self-energy} can be expressed as~\cite{Ko:1992tp, Cobos-Martinez:2017vtr}
\begin{align}
\Pi^{\mu\nu}_{\rm loop}(p^2) =&~ 2ig_\phi^2\int \frac{\dd^4q}{(2\pi)^4} 
\frac{S^{\mu\nu}_{\rm loop}}{D_K(q) D_{\bar{K}}(p-q)},\\
\Pi^{\mu\nu}_{\rm con}(p^2) =&~ 2ig_\phi^2\int \frac{\dd^4q}{(2\pi)^4} 
\frac{S^{\mu\nu}_{\rm con}}{D_K(q)},
\end{align}
where $q$ is the kaon four-momentum and the factor two is coming from the charged and neutral $K\bar{K}$ loops~\cite{Cobos-Martinez:2017vtr}.
Here, we assume that the kaon is a stable particle and has no width.
The numerators are given by 
\begin{align}
    S_{\rm loop}^{\mu\nu}=(2q-p)^\mu(2q-p)^\nu, \qquad 
    S_{\rm con}^{\mu\nu}=-2g^{\mu\nu},
\end{align} 
while the kaon and antikaon propagators are expressed by
\begin{align}
    D_K(q) &= q^2 - m_K^{2} + i\epsilon,\\
    D_{\bar{K}}(p-q) &= (p - q)^2 - m_K^{2} + i\epsilon,
\end{align}
where $m_K(=m_{\bar{K}})$ is the kaon scalar mass.
Projecting onto the transverse and longitudinal polarization modes ($\lambda=T,L$), we obtain
\begin{align}
\Pi^{\lambda}_\mathrm{total}(p^2) =&~ 2ig_\phi^2\int \frac{\dd^4q}{(2\pi)^4}\left( 
\frac{\mathcal{O}^{\lambda}_{\rm loop}}{D_K D_{\bar{K}}}+\frac{\mathcal{O}^{\lambda}_{\rm con}}{D_K}\right),
\end{align}
where the operator $\mathcal{O}_{\lambda}$ is defined by
\begin{align}\label{eq:operator}
    \mathcal{O}_{L} = P^L_{\mu\nu}S^{\mu\nu},\qquad \mathcal{O}_{T} = \frac{1}{2}P^T_{\mu\nu}S^{\mu\nu}.
\end{align}

\subsection{In-medium modifications}

In nuclear matter, the self-energy of propagating $\phi$ meson can be written as~\cite{Cobos-Martinez:2017vtr} 
\begin{align}\label{eq:med-selfs}
       \Pi_\mathrm{total}^{\lambda}(m_\phi^{*2},\bm{p}^2) = 2ig_\phi^2 \int \frac{\dd^4q}{(2\pi)^4} 
       \left( \frac{\mathcal{O}_{\rm loop}^{\lambda*}}{D_K^*D_{\bar{K}}^*} + \frac{\mathcal{O}_{\rm con}^{\lambda*}}{D_K^*} \right).
\end{align}
Since Lorentz symmetry is broken in nuclear matter, the self-energy depends on two independent variables, $m_\phi^*$ and $\bm{p}$.\footnote{More precisely, the self-energy depends on the pole energy $E_\phi^*$ and the three-momentum $\bm{p}$. However, using the mass-shell relation $m_\phi^{*2}=E_\phi^{*2}-\bm{p}^2$, it can equivalently be expressed in terms of $m_\phi^*$ and $\bm{p}$}.
The momentum-dependent pole mass $m_\phi^*(\bm{p})$ is determined self-consistently from the mass equation in Eq.~\eqref{eq:phi_massx}.
The kaon and antikaon propagators are given by
\begin{align}
   D_K^*=&~ (q^{0*})^2 - \bm{q}^2-m_K^{*2}+i\epsilon,\\
   D_{\bar{K}}^* =&~ (p^{0*}-q^{0*})^2-(\bm{p}-\bm{q})^2-m_K^{*2}+i\epsilon,
\end{align}
Here, the scalar mean field affects the kaon mass, 
\begin{equation}
    m_K^* \equiv m_K - V_\sigma^K,
\end{equation}
while the vector mean field modifies the kaon energy, modifying the dispersion relation
\begin{equation}
    q^{0*} = q^0 + V_\omega^K, \qquad   (p-q)^{0*} = (p-q)^0 - V_\omega^K.
\end{equation}
Note that we denote all in-medium quantities with an asterisk. 

To evaluate the integral, we can do a variable shift,
\begin{align}\label{eq:shift1}
    q^0  \to q^{0} - V_\omega^K,
\end{align}
and can obtain the same formal expression as in Eq.~\eqref{eq:med-selfs}, 
with operators in the numerator shifted as
\begin{align}
{\mathcal{O}}_{\rm loop}^{\lambda*} \;\to\;
{\mathcal{O}}_{\rm loop}^{\lambda*}(p, q^{0}-V_\omega^K).
\end{align}
Note that, the operator ${\mathcal{O}}_{\rm loop}^{\lambda*}$ can depend explicitly on $q^0$, and this dependence 
is modified by the vector mean field $V_{\omega}$. 
As a consequence, the transverse and longitudinal polarization modes become separated.
However, ${\mathcal{O}}_{\rm con}^{\lambda*}$ does not depend on $q^0$ and therefore is not affected by $V_{\omega}^K$.
After the shift in Eq.~\eqref{eq:shift1}, the in-medium propagators can be defined without $V_\omega^K$ dependence as
\begin{align}
D_K^* &= (q^{0})^2 - \bm{q}^2 - m_K^{*2} + i\epsilon,\\
D_{\bar{K}}^* &= (p^{0}-q^{0})^2 - (\bm{p}-\bm{q})^2 - m_K^{*2} + i\epsilon,
\end{align}
and this notation will be used throughout this work.

\subsection{In-medium $\phi$ mass and spectral functions}

Once the $\phi$ self-energy $\Pi(p)$ is obtained, the respective in-medium mass and decay width can be determined. When the kinematic condition $m_\phi^* > 2m_K^*$ is satisfied, the self-energy develops an imaginary part, corresponding to the opening of the $\phi \to K\bar{K}$ decay channel in the medium. 

The in-medium mass of the $\phi$ meson is then determined from the real part of the self-energy through
\begin{eqnarray}\label{eq:phi_massx}
m_\phi^{*2} = (m_{\phi}^{0})^2 + \text{Re}~\Pi(m_\phi^{*2},\bm{p}^2),
\end{eqnarray}
which determines the momentum-dependent pole mass
$m_\phi^*(\bm p)$ self-consistently for a given value of
$\bm p^2$.
Here, $m_{\phi}^0$ is the bare $\phi$-meson mass parameter determined in vacuum.
The corresponding intrinsic in-medium decay width for the $\phi \to K\bar{K}$ process is obtained from the imaginary part of the self-energy,
\begin{equation}\label{eq:phi_widthx}
\Gamma_\phi^{\ast} = -\frac{1}{m_\phi^\ast}\text{Im}~\Pi(m_\phi^{\ast 2},\bm{p}^2).
\end{equation}
This quantity characterizes the medium-induced broadening of the $\phi$ meson associated with its self-energy.  
For a propagating $\phi$ meson, the decay width observed in the laboratory frame is reduced due to Lorentz time dilation. Introducing $\gamma = E_\phi^\ast/m_\phi^\ast$, the corresponding laboratory-frame width is given by $\Gamma_{\phi,\mathrm{lab}}^\ast = \Gamma_\phi^{\ast}/\gamma$.
In Sec.~\ref{sec:regularization}, the self-energy is evaluated using two different regularization schemes in order to assess the robustness and model dependence of the results. 

To make connection with observables in the invariant-mass distribution, it is useful to introduce the in-medium $\phi$-meson spectral function using a Breit-Wigner parametrization~\cite{Kim:2019ybi}.
\begin{align}
A_\phi^\lambda(s,\bm{p}^2)
= \frac{C_\lambda}{\pi}
\frac{ \sqrt{s}\Gamma_\phi^{*}(\bm{p})}
{(s - m_\phi^{*2}(\bm{p}))^2 + s\Gamma_\phi^{*2}(\bm{p})},
\end{align}
where the intrinsic decay width $\Gamma_\phi^{*}$ is used.
The normalization constant $C_\lambda$ is determined by the condition
\begin{align}
\int_0^\infty \dd s\; A_\phi^\lambda(s,\bm{p}^2) = 1.
\end{align}
Since most dilepton measurements have so far not been able to resolve individual polarization states, 
it is natural to introduce a polarization-averaged spectral function, defined as
\begin{align}
A_\phi^{\mathrm{ave}}(s,\bm{p}^2)
= \frac{1}{3}
\biggl[
A_\phi^L(s,\bm{p}^2)
+ 2 A_\phi^T(s,\bm{p}^2)
\biggr] .
\end{align}
This polarization averaging provides a practical connection between theoretical predictions and dilepton measurements.

\subsection{In-medium kaon mass}
\label{sec:kaon-mass}

The modification of the kaon mass and energy in nuclear matter is an important input in our study.
The kaon mass and energy in medium are evaluated in the QMC model~\cite{Tsushima:1997df}, which was proposed by Guichon~\cite{Guichon:1987jp} and has been successfully applied to describe both infinite nuclear matter and finite nuclei (see Refs.~\cite{Saito:2005rv,Guichon:2018uew} for reviews). 
Here we summarize only the elements relevant to the present work.

In the QMC model, nuclear matter is treated in its rest frame within the mean-field (Hartree) approximation, where the scalar and vector meson fields are spatially constant. 
The light quarks $q=(u, d)$ inside a hadron bag couple directly to these mean fields, while the strange quark remains unaffected. Here we assume SU(2) symmetry for the light quarks ($m_q=m_u=m_d$).
The corresponding Dirac equations for the quark fields in symmetric nuclear matter,
\begin{align}
(i\slashed{\partial} - m_q^* \mp \gamma^0 V_\omega^q) \begin{pmatrix}
        \psi_q \\ \psi_{\bar{q}}
    \end{pmatrix} = 0, \quad
(i\slashed{\partial} - m_s) \begin{pmatrix}
        \psi_s \\ \psi_{\bar{s}}
    \end{pmatrix} = 0,
\end{align}
lead to an effective light-quark mass
\begin{align}
    m_q^* \equiv m_q - V_\sigma^q.
\end{align}
Since we restrict our study to symmetric nuclear matter, the contribution from the $\rho$ meson mean field can be neglected. 
The mean-field potentials are then written as
\begin{equation}
    V_\sigma^q = g_\sigma^q \sigma, \qquad 
    V_\omega^q = g_\omega^q \omega ,
\end{equation}
where $g_\sigma^q$ and $g_\omega^q$ are the corresponding quark--meson coupling constants.

The normalized static solution for a ground-state quark or antiquark of flavor $f$ in a kaon $K$ can be expressed as
\begin{equation}
    \psi_f(x) = N_f\, e^{-i\epsilon_f t/R_K^*}\, \psi_f(\bm{r}),
\end{equation}
where $N_f$ is the normalization factor and $\psi_f(\bm{r})$ denotes the spatial and spinor wave function.  
The in-medium bag radius, $R_K^*$, is determined by imposing the stability condition on the hadron mass with respect to variations of the radius.  

The corresponding eigenenergies (in units of $1/R_K^*$) take the form
\begin{align}
    \begin{pmatrix}
        \epsilon_u \\ \epsilon_{\bar{u}}
    \end{pmatrix} = \Omega_q^* \pm R_K^* V_\omega^q, \qquad
    \begin{pmatrix}
        \epsilon_d \\ \epsilon_{\bar{d}}
    \end{pmatrix} = \Omega_q^* \pm R_K^* V_\omega^q,
\end{align}
and $\epsilon_s = \epsilon_{\bar{s}} = \Omega_s.$
Then, the excitation energy of $K$ and $\bar{K}$ mesons with zero momenta are given by
\begin{align}
       \begin{pmatrix}
        E_{K^+} \\ E_{K^-}
    \end{pmatrix} = m_K^* \pm V_\omega^K, \qquad 
       \begin{pmatrix}
        E_{K^0} \\ E_{\bar{K}^0}
    \end{pmatrix} = m_K^* \pm V_\omega^K.
\end{align}
In a constant vector field, the Lorentz invariance of the dispersion relation for the $K$ and $\bar{K}$ mesons moving with momentum $\bm{q}$ are given by~\cite{Brown:1992ve,Tsushima:1997df}
\begin{align}\label{eq:kaon_energy}
         \begin{pmatrix}
        E_{K^+}(\bm{q}) \\ E_{K^-}(\bm{q})
    \end{pmatrix} &= \sqrt{m_K^{*2} + \bm{q}^2} \pm V_\omega^K, \\ 
       \begin{pmatrix}
        E_{K^0}(\bm{q}) \\ E_{\bar{K}^0}(\bm{q})
    \end{pmatrix} &= \sqrt{m_K^{*2} + \bm{q}^2} \pm V_\omega^K,
\end{align}
which is useful for the calculation of the in-medium $K\bar{K}$ loop.
Here, the background nuclear matter is treated as static, uniform, and at rest.
In such a medium the vector field is generated by nucleons that are, on average, at rest, so that the expectation value of the vector mean field remains purely time-like. 

The in-medium kaon mass is then calculated as
\begin{equation}
    m_K^* = \sum_{j=q,\bar{s}} \frac{n_j\Omega_j^* - z_K}{R_K^*} + \frac{4\pi}{3} R_K^{*3} B,
\end{equation}
where $z_K$ accounts for the center-of-mass and gluon fluctuation corrections, and $B$ is the bag constant. 
The in-medium mass is obtained by imposing the stability condition
\begin{equation}
    \frac{\dd m_K^*}{\dd R_K^*} = 0.
\end{equation}
Here, the quark eigenenergies are
\begin{align}
    \Omega_q^* &= \Omega_{\bar{q}}^* = \sqrt{x_q^2 + (R_K^* m_q^*)^2}, \\
    \Omega_s^* &= \Omega_{\bar{s}} = \sqrt{x_s^2 + (R_K^* m_s^*)^2},
\end{align}
where $x_{q,s}$ are the lowest bag eigenfrequencies, and $n_q (n_{\bar{q}})$ denote the numbers of quarks (antiquarks) of flavors $q$ and $s$. After a self-consistent calculation, the in-medium kaon mass can be parameterized as
\begin{equation}
    m_K^* \equiv m_K - V_\sigma.
\end{equation}
We note that this kaon mass remains unchanged in the moving kaon frame due to its scalar nature.

\section{Regularization Schemes}
\label{sec:regularization}

In this work, we consider two different regularization schemes to evaluate the $\phi$-meson self-energy.
We do not consider the three-momentum form factor employed for the zero-momentum case in the previous work of Ref.~\cite{Cobos-Martinez:2017vtr}, because this procedure explicitly breaks Lorentz
covariance, and is hence not suitable for the finite-momentum case, which is the focus of the present study.
To overcome this limitation, we employ a covariant form factor as well as dimensional regularization~\cite{Ko:1992tp}. 
Using two different schemes allows us to check the scheme dependence and assess the robustness of our results.

\subsection{Form factor regularization}

We first consider a covariant form factor with a multipole Ansatz to regulate the self-energy of the $\phi$ meson shown in Fig.~\ref{fig:self-energy}. 
The total self-energy is given by
\begin{align}
& \Pi_\mathrm{total}^\lambda(m_\phi^{*2},\bm{p}^2)\nonumber\\
&~=2ig_\phi^2 \int \frac{\dd^4 q}{(2\pi)^4} \left(\frac{\mathcal{O}_{\rm loop}^{\lambda*} }{D_K^* D_{\bar{K}}^*} + \frac{\mathcal{O}_{\rm con}^{\lambda*} }{D_K^*}\right)F(q,\Lambda) \nonumber\\
        &~ =2ig_\phi^2 (\Lambda^2 -m_K^{*2})^2\nonumber\\
        &~~~~\times \int \frac{\dd^4 q}{(2\pi)^4} \left(\frac{\mathcal{O}_{\rm loop}^{\lambda*}}{D_\Lambda^2D_K^*D_{\bar{K}}^*} +  \frac{\mathcal{O}_{\rm con}^{\lambda*} }{D_\Lambda^2D_K^*}\right), \label{eq:se-ff}
\end{align}
with $D_\Lambda= q^2 - \Lambda^2 + i\epsilon$. 
The regulator is chosen as
\begin{align}
    F(q, \Lambda) = \frac{(\Lambda^2 -m_K^{*2})^2}{(q^2 - \Lambda^2 + i\epsilon)^2},
\end{align}
where the cutoff parameter $\Lambda$ is assumed to be independent of the nuclear density.
As mentioned before, the total contribution can be decomposed into loop and contact terms,
$\Pi^\lambda_{\rm total}=\Pi^\lambda_{\rm loop}+\Pi^\lambda_{\rm con}$.

The operators appearing in the numerator, defined in Eq.~\eqref{eq:operator}, are evaluated as
\begin{align}
    \mathcal{O}_{\rm loop}^{T*} =&~\frac{1}{2}P_{\mu\nu}^T(2q-p)^\mu(2q-p)^\nu =2 \bm{q}_\perp^2, \label{eq:opt-t}\\ 
    \mathcal{O}_{\rm loop}^{L*} =&~P_{\mu\nu}^L(2q-p)^\mu(2q-p)^\nu \nonumber\\
                =&~\frac{4}{m_\phi^{*2}}[(q^0-V_\omega^K)|\bm{p}|-q^z E_\phi^*]^2,\label{eq:opt-l}\\ 
    \mathcal{O}_{\rm con}^{L(T)*} =&~\frac{1}{2}P_{\mu\nu}^T (-2g^{\mu\nu})= P_{\mu\nu}^L (-2g^{\mu\nu})= 2,\label{eq:opt-con}
\end{align} 
where a variable shift of $q^0$, given in Eq.~\eqref{eq:shift1}, has been performed.
The operators for the transverse and longitudinal modes differ in structure. 
For the transverse mode, the operator does not depend on the $\phi$-meson three-momentum $|\bm{p}|$, whereas for the longitudinal mode it exhibits an explicit momentum dependence, together with the vector mean field $V_\omega^K$. 
The absence of momentum dependence in the transverse operator follows from $\bm{p}_\perp=0$, such that only the longitudinal momentum component is nonvanishing.
By contrast, the operator for the contact term is identical for both longitudinal and transverse projectors.

Physically, this difference originates from the tensor structure of the derivative $\phi K\bar K$ interaction in Eq.~\eqref{eq:lagrangian1}, which gives rise to the momentum-dependent vertex factor $(2q-p)^\mu$. When projected onto the transverse and longitudinal polarization states, this vertex factor generates the different operator structures shown in Eqs.~\eqref{eq:opt-t} and \eqref{eq:opt-l}. The transverse projector selects components perpendicular to the $\phi$-meson momentum and therefore removes any explicit dependence on $|\bm p|$ from the loop operator. By contrast, the longitudinal projector couples the loop energy and longitudinal momentum, allowing both the vector mean field $V_\omega^K$ and the $\phi$-meson momentum to enter the self-energy. Consequently, only the longitudinal self-energy acquires a momentum-dependent medium modification, while the transverse self-energy remains momentum independent within the present calculation.

After introducing the Feynman parameterization, Wick rotation, and Euclidean integration, the total self-energies for the transverse and longitudinal modes are given by
\begin{align}
   \Pi_\mathrm{total}^T(m_\phi^{*2},\bm{p}^2) =&~ N \int_0^1 \dd x \Biggl[ \int_0^{1 - x} \dd y~\frac{(1 - x - y)}{C} - \frac{x}{E} \Biggr], \qquad \label{eq:ffT}\\
   \Pi_\mathrm{total}^L(m_\phi^{*2},\bm{p}^2) =&~ N \int_0^1 \dd x \Biggl[ \int_0^{1 - x} \dd y~(1 - x - y) \nonumber\\
        &~ \times \left(\frac{1}{C} - \frac{2|\bm{p}|^2 (V_\omega^K)^2}{m_\phi^{*2}}\frac{1}{C^2}\right) - \frac{x}{E} \Biggr], \label{eq:ffL}
\end{align}
where $N={g_\phi^2(\Lambda^2 -m_K^{*2})^2 }/{4\pi^2}$, with the denominator factors given by
\begin{align}
    C =&~ x(1-x)m_\phi^{*2} - (x+y) m_{K}^{*2} - (1-x-y)\Lambda^2, \\
    E =&~ -x\Lambda^2 - (1-x) m_K^{*2}.
\end{align}
Note that the contact term corresponds to the contribution proportional to $E$, while the loop term involves $C$.
To compute the real part, the integrals are evaluated using the principal-value calculation, where the numerical implementation avoids singularities by smearing, for example through a Breit–Wigner form,
\begin{align}
    \frac{1}{C} \to \frac{C}{C^2 + \Gamma^2}, \hspace{1cm} \frac{1}{C^2} \to \frac{C^2-\Gamma^2}{(C^2 + \Gamma^2)^2},
\end{align}
where $\Gamma$ is a small regulator parameter.
An explicit analytical calculation is also possible.
The details of calculation and analytical expressions for these integrals are provided in App.~\ref{app:ff}.

While the real part depends on the regulator $\Lambda$, the imaginary part is given by
\begin{align}
    \mathrm{Im}~\Pi^{T}_\mathrm{total}(m_\phi^{*2},\bm{p}^2) =&~ - \frac{g_\phi^2}{24\pi} m_\phi^{*2} \beta^3, \\
    \mathrm{Im}~\Pi^{L}_\mathrm{total}(m_\phi^{*2},\bm{p}^2) =&~ - \frac{g_\phi^2}{24\pi} m_\phi^{*2} \beta^3 \left( 1 + \frac{12}{\beta^2} \frac{ (V_\omega^K)^2 |\bm{p}|^2}{m_\phi^{*4}}\right),
\end{align}
with $\beta=\sqrt{1-4m_K^{*2}/m_\phi^{*2}}$.
The imaginary part arises solely from the loop diagram, whereas the contact term contributes only to the real part. 
Putting into Eq.~\eqref{eq:phi_widthx}, the intrinsic decay width can be explicitly expressed by
\begin{align}
    \Gamma^{*T}_\phi(m_\phi^{*2},\bm{p}^2) =&~ \frac{g_\phi^2}{24\pi} m_\phi^{*} \beta^3, \\
    \Gamma^{*L}_\phi(m_\phi^{*2},\bm{p}^2) =&~ \frac{g_\phi^2}{24\pi} m_\phi^{*} \beta^3 \left( 1 + \frac{12}{\beta^2} \frac{ (V_\omega^K)^2 |\bm{p}|^2}{m_\phi^{*4}}\right). \label{eq:width-p2}
\end{align}

\subsection{Dimensional Regularization}

We now turn to dimensional regularization, which was used for instance in Ref.~\cite{Ko:1992tp}.
The $\phi$-meson self-energy can be written as 
\begin{align}\label{eq:med-self}
       \Pi_\mathrm{total}^{\lambda}(m_\phi^{*2},\bm{p}^2) =&~ 2ig_\phi^2 \int \frac{\dd^4q}{(2\pi)^4} 
       \left( \frac{\mathcal{O}_{\rm loop}^{\lambda*}}{D_K^*D_{\bar{K}}^*} + \frac{\mathcal{O}_{\rm con}^{\lambda*}}{D_K^*} \right),
\end{align}
without introducing an explicit form factor.

For the contact term, the self-energy is identical for transverse and longitudinal polarizations and is given by
\begin{align}
\Pi_\mathrm{con}^{\lambda}(m_\phi^{*2},\bm{p}^2) = 4ig_\phi^2 \int \frac{\dd^4 q}{(2\pi)^4} \frac{1}{D_K^*},
\end{align}
where we have used $\mathcal{O}_{\rm con}^{\lambda*}=2$, as in Eq.~\eqref{eq:opt-con}.
After applying the Feynman parameterization and momentum shift to complete the square, the self-energy for the loop contribution can be expressed as
\begin{align}
    \Pi^{\lambda}_{\rm loop}(m_\phi^{*2},\bm{p}^2)
    = 2ig_\phi^2 \int_0^1 \dd x \int \frac{\dd^4 q}{(2\pi)^4}
    \frac{{\mathcal{O}}^{\lambda*}_{\rm loop}}{D_\Delta^{*2}},
\end{align}
where $D_\Delta^*={q}^{2} - \Delta^* + i\epsilon$ and $\Delta^* = m_K^{*2} - x(1-x)m_\phi^{*2}$.
The explicit forms of the operators for the transverse and longitudinal modes are the same as in Eqs.~\eqref{eq:opt-t} and ~\eqref{eq:opt-l}, respectively.

Including both the loop and contact contributions, the transverse self-energy can be written as
\begin{align}
    \Pi_\mathrm{total}^T(m_\phi^{*2},\bm{p}^2) 
    =&~ 4ig_\phi^2 \int \frac{\dd^d {q}}{(2\pi)^d}  \biggl( \int_0^1 \dd x 
       \frac{P_{\mu\nu}^T {q}^{\mu}{q}^{\nu}}{D_\Delta^{*2}} + \frac{1}{D_K^*} \biggr),
\end{align}
where, in $d$ dimensions, we have used the identity $\bm{q}_\perp^{2} = P_{\mu\nu}^Tq^\mu q^\nu$.
The momentum integrals are evaluated using dimensional regularization in $d = 4 - \epsilon$ dimensions.
Keeping only finite terms, the renormalized transverse self-energy reads
\begin{align}\label{eq:repi_T}
&\Pi_\mathrm{total}^T(m_\phi^{*2},\bm{p}^2)~\nonumber\\
    &~=-\frac{g_\phi^2}{4\pi^2}
    \int_0^1 \dd x\, \biggr(\Delta^*\ln\frac{\Delta^*}{\mu^2} - m_K^{*2} \ln \frac{m_K^{*2}}{\mu^2} - a(\mu)\frac{m_\phi^2}{6} \biggr),
\end{align}
where the first two terms arise from the loop and contact contributions, respectively.
Here $\mu$ denotes the renormalization scale, and $a(\mu)$ is a dimensionless subtraction constant that absorbs the ultraviolet divergence.
The medium-independent factor $m_\phi^2/6$ originates from the combined loop and contact contributions.

For the longitudinal mode, the loop contribution is given by
\begin{align}
&\Pi_\mathrm{loop}^L(m_\phi^{*2},\bm{p}^2)\nonumber\\
&~~= \frac{8ig_\phi^2}{m_\phi^{*2}} \int_0^1 \dd x \int \frac{\dd^4 {q}}{(2\pi)^4} \frac{\left( [{q}^0 - V_\omega^K] |\bm{p}| - {q}^z E_\phi^* \right)^2}{ D_\Delta^{*2}}.
\end{align}
By adding the contact contribution and following procedures similar to those used for the transverse mode, we obtain
\begin{align}\label{eq:repi_L}
& \Pi_\mathrm{total}^L(m_\phi^{*2},\bm{p}^2) \nonumber\\
&~ =-\frac{g_\phi^2}{4\pi^2} \int_0^1 \dd x  \Biggl[ \Delta^* \ln\frac{\Delta^*}{\mu^2} - m_K^{*2} \ln \frac{m_K^{*2}}{\mu^2}  \nonumber\\
&~~~~ - 2 (V_\omega^K)^2\frac{|\bm{p}|^2}{m_\phi^{*2}} \left( \ln\frac{\Delta^*}{\mu^2} + 1 \right) 
    - a(\mu)\frac{m_\phi^2}{6} \Biggr].
\end{align}
The details of calculations and analytic expressions are provided in App.~\ref{app:dim-reg}.

While the real part depends on the renormalization scale $\mu$ and the subtraction constant $a(\mu)$, the imaginary parts of the transverse and longitudinal self-energies are again independent of the regularization scheme and are given by
\begin{align}
    \mathrm{Im}~\Pi^{T}_\mathrm{total}(m_\phi^{*2},\bm{p}^2) =&~ - \frac{g_\phi^2}{24\pi} m_\phi^{*2} \beta^3, \\
    \mathrm{Im}~\Pi^{L}_\mathrm{total}(m_\phi^{*2},\bm{p}^2) =&~ - \frac{g_\phi^2}{24\pi} m_\phi^{*2} \beta^3 \left( 1 + \frac{12}{\beta^2} \frac{ (V_\omega^K)^2 |\bm{p}|^2}{m_\phi^{*4}}\right).\qquad 
\end{align}
These expressions agree with those obtained using the covariant form-factor regularization.

\section{Model parameters}
\label{sec:parameter}

In this section, we discuss how the parameters of our model are determined. 
Important inputs are the in-medium kaon properties, which are obtained from the QMC model~\cite{Tsushima:1997df}. 
The parameters in the self-energy terms of the effective Lagrangian are fixed using the relevant decay widths~\cite{ParticleDataGroup:2024cfk}. 
Moreover, each regularization scheme introduces additional parameters that must also be determined.

\subsection{Self-energy parameters}

\begin{figure}[b]
    \centering
    \includegraphics[width=0.42\textwidth]{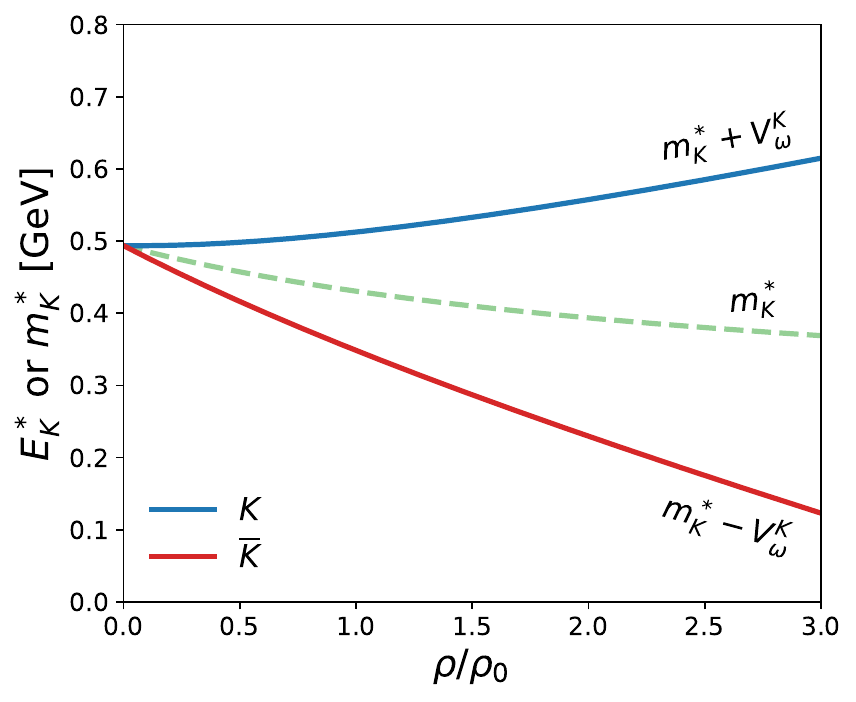} 
    \caption{In-medium mass and energy of a kaon at rest in nuclear matter calculated using the QMC model~\cite{Tsushima:1997df}. The kaon mass shown with green dashed line is reduced by the scalar mean field $V_\sigma^K$, while the $K\bar{K}$ energy splitting arises from the vector mean field. The kaon energy exhibits a slight repulsion, consistent with experimental observations. } 

    \label{fig:kaon-mass}
\end{figure}

First, we fix the model parameters entering the $\phi$-meson self-energy. The vacuum masses of the $\phi$ meson and kaon are taken as $m_\phi^{\mathrm{expt.}} = 1019.461~\mathrm{MeV}$ and $m_K^{\mathrm{expt.}} = 493.7~\mathrm{MeV}$, respectively. The $\phi KK$ coupling constant is then determined as $g_\phi = 4.517$ by reproducing the decay width of the $\phi$ meson in vacuum at rest, $\Gamma_\phi^{\mathrm{expt.}} = 4.249~\mathrm{MeV}$.

\begin{figure*}[t]
    \centering
    \includegraphics[width=0.42\textwidth]{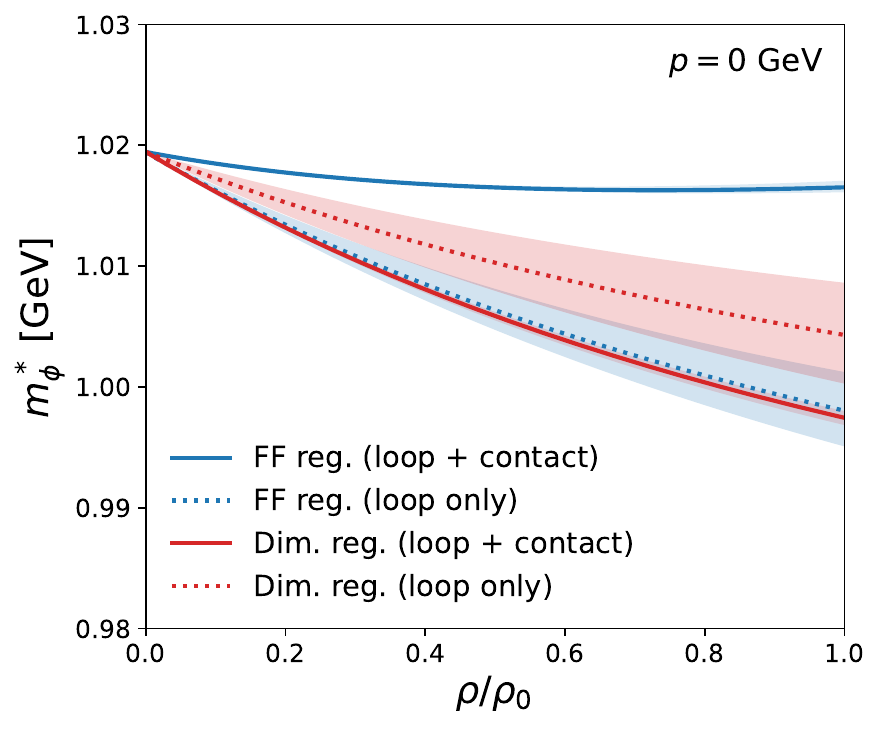}
    \includegraphics[width=0.42\textwidth]{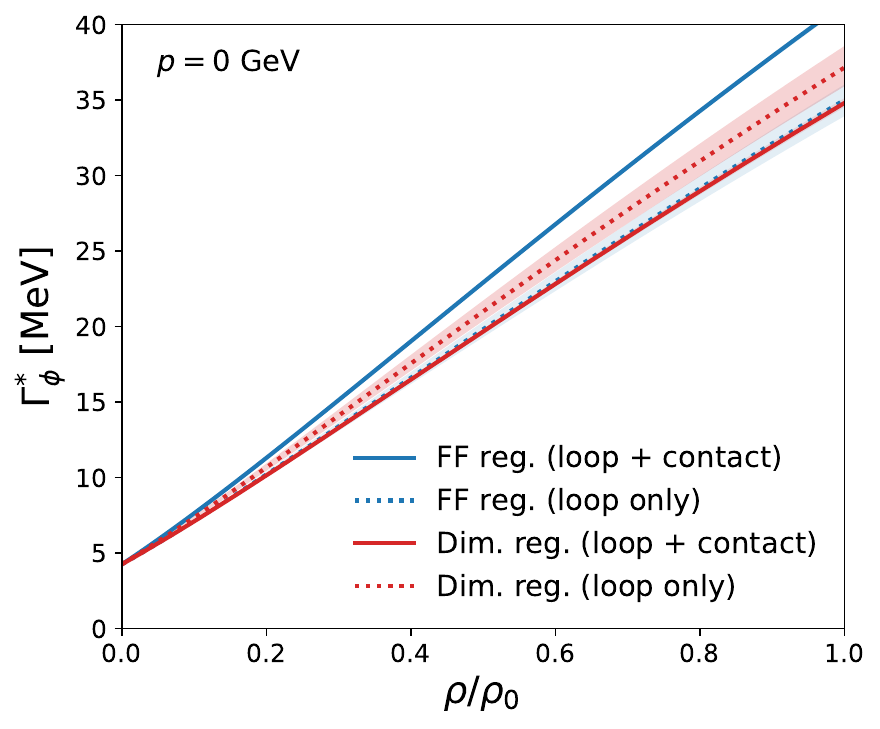}
    \caption{
    In-medium mass and width of the $\phi$ meson at rest as functions of nuclear density, computed using two different regularization schemes. The solid (dotted) lines correspond to calculations with (without) the contact term.
    }
    \label{fig:property-rest}
\end{figure*}

Next, we briefly review the parameters used in the QMC model for the in-medium kaon properties~\cite{Tsushima:1997df}. We use the current quark masses $(m_q, m_s) = (5, 250)~\mathrm{MeV}$, with vacuum bag radii $R_N = 0.8~\mathrm{fm}$ and $R_K = 0.574~\mathrm{fm}$ for the nucleon and kaon, respectively. The zero-point parameters $z_{N,K} = 3.295$ (the same values within the digits) and the bag constant $B^{1/4} = 170~\mathrm{MeV}$ are taken to be identical for both hadrons, chosen to reproduce their vacuum properties.
In symmetric nuclear matter, the quark-meson coupling constants are fixed to reproduce the saturation point, with a binding energy of 15.7~MeV at $\rho_0 = 0.15~\mathrm{fm}^{-3}$, yielding $(g_\sigma^q, g_\omega^q) = (5.69, 2.72)$. 

The resulting in-medium kaon energy and effective mass at rest are displayed in Fig.~\ref{fig:kaon-mass}. The Lorentz-scalar kaon mass decreases with increasing density due to the attractive scalar mean field $V_\sigma^K$, as indicated by the green dashed line. In contrast, the vector mean field, $V_\omega^K = g_{K\omega}^q \omega$, generates a splitting between kaon and antikaon energies. To account for the slight repulsion observed in the $K^+N$ interaction, of order 20~MeV~\cite{Fuchs:2005zg}, the quark–vector-meson coupling is phenomenologically enhanced to $g_{K\omega}^q = 1.4^2 g_\omega^q$. 
We note that alternative mechanisms~\cite{Arifi:2023jfe} could also produce a similar repulsive effect without modifying the coupling strength~$g_\omega^q$. 
At normal nuclear matter density, the kaon effective mass is reduced by approximately $13\%$~\cite{Tsushima:1997df}, slightly larger than reported in Ref.~\cite{Ko:1992tp}.

\subsection{Regularization parameters}

Lastly, we will describe how the parameters are fixed in both regularization schemes. 

\paragraph{Form factor regularization}  
In the covariant form factor scheme, two parameters must be fixed: the cutoff $\Lambda$ and the bare mass $m_\phi^0$ in Eq.~\eqref{eq:phi_massx}.  
To illustrate the model uncertainty, we consider $\Lambda = 0.8, 0.9,$ and $1.0~\text{GeV}$, with the corresponding bare masses $m_\phi^0 = 1.015, 1.005,$ and $0.990~\text{GeV}$ chosen to reproduce the physical $\phi$ mass in vacuum.  
These cutoffs are smaller than in the three-momentum form factor scheme~\cite{Cobos-Martinez:2017vtr} because the covariant form factor regulates both $\bm{q}$ and $q^0$, providing stronger suppression at large momenta. We note that, when only the loop diagram is considered, the bare masses become $m_\phi^0 = 1.055$, $1.073$, and $1.093$ GeV, respectively.

\paragraph{Dimensional regularization}

In this scheme, two parameters must be fixed: the renormalization scale $\mu$ and the subtraction constant $a(\mu)$. The bare mass $m_\phi^0$ is set to the physical $\phi$ mass in vacuum, and the parameters are determined by requiring $\mathrm{Re}~\Pi = 0$ at vacuum.  
Unlike the form factor scheme, variations in $m_\phi^0$ can effectively be absorbed into $a(\mu)$.  
Instead of fixing $\mu$ to the kaon mass~\cite{Ko:1992tp}, we explore several values of $\mu = 0.5, 0.6,$ and $0.7~\text{GeV}$, with corresponding 
$a(\mu) = -0.1920, 0.1727,$ and $0.4810$, different from the previous work~\cite{Ko:1992tp}  because a constant term is retained rather than fully absorbed into $a(\mu)$ (See App.~\ref{app:dim-reg} for details). We note that, when only the loop diagram is considered, the subtraction constants become $a(\mu) = 0.4606$, $0.7610$, and $1.0150$ GeV, respectively.

We emphasize that, for a fixed choice of the regularization scale $\mu$
(or cutoff $\Lambda$), the same bare $\phi$-meson mass and the same
renormalization condition are used for both longitudinal and transverse
polarization modes.

\section{Results and Discussion} 
\label{sec:result}

In this section, we present the numerical results on the in-medium behavior of the $\phi$ meson.
We first examine medium-induced modifications to its mass and width at rest using two different regularization schemes.
We then extend the analysis to the case of a $\phi$ meson propagating with finite momentum, 
which constitutes the main focus of the present work.
Finally, we discuss the resulting unpolarized spectral functions and their implications for experiments~\cite{YokkaichiE16Proposal, Naruki:2012eka, Aoki:2023qgl, Aoki:2024ood, Sako:2024oxb}.

\subsection{In-medium properties at rest}

We first examine the in-medium $\phi$-meson mass and width at rest for densities up to normal nuclear density $\rho_0$, relevant to experiments such as those at J-PARC. At zero momentum, the longitudinal and transverse modes are degenerate. Figure~\ref{fig:property-rest} displays the density dependence of the in-medium mass $m_\phi^*$ and width $\Gamma_\phi^*$ obtained with both the form-factor and dimensional-regularization schemes. The results are presented by including the loop contribution with or without the contact term.

In Fig.~\ref{fig:property-rest}, we find that the loop contribution alone gives similar results in both regularization schemes. The inclusion of the contact term leads to a significantly weaker mass reduction in the form-factor regularization, whereas in dimensional regularization the mass decreases only slightly further. However, for sufficiently large values of the renormalization scale $\mu$, dimensional regularization can instead predict a weaker mass reduction.
Quantitatively, at $\rho_0$, the loop-only contribution yields a mass reduction of approximately 10--20 MeV. After including the contact term, the mass reduction is about 20 MeV in dimensional regularization, whereas it is reduced to only a few MeV in the form-factor regularization. In addition, the decay width is broadened by approximately $35$--$40~\mathrm{MeV}$. Overall, the predicted mass shift and substantial broadening are consistent with the recent reanalysis~\cite{KEK-PSE325:2025fms} of KEK-E325 dilepton data~\cite{KEK-PS-E325:2005wbm}. 

Although the two regularization schemes exhibit quantitatively different behavior after the contact term is included, the quantitative results remain broadly consistent with previous studies in each scheme.
For instance, in the three-dimensional form-factor regularization, Ref.~\cite{Cobos-Martinez:2017vtr} showed that including the contact term weakens the mass shift, requiring a larger cutoff to reproduce the mass shift obtained from the loop-only calculation. For this reason, calculations employing form-factor regularization often consider only the loop contribution and omit the contact term. In contrast, our results obtained with dimensional regularization are consistent with those of Ko \textit{et al}.~\cite{Ko:1992tp}.

Despite quantitative differences, both regularization schemes predict qualitatively similar medium modifications of the $\phi$ meson, namely a downward mass shift and a substantial broadening in nuclear matter. Since the form-factor and dimensional-regularization approaches correspond to different treatments of the ultraviolet dynamics, an exact correspondence between $\Lambda$ and $\mu$ is not expected. We therefore regard the spread of predictions obtained from the two schemes as an estimate of the theoretical uncertainty related to the regularization procedure.

Furthermore, the error bands in Fig.~\ref{fig:property-rest} represent the dependence on the cutoff parameter $\Lambda=0.8$--1.0 GeV in the form-factor scheme and the renormalization scale $\mu=0.5$--0.7 GeV in dimensional regularization.  We find that the loop contribution alone exhibits a strong dependence on the cutoff or renormalization scale in both regularization schemes. However, the inclusion of the contact term significantly reduces this dependence by partially compensating for the regulator-dependent part of the loop contribution, leading to a self-energy that is less sensitive to the cutoff/scale parameter. Consequently, the remaining parameter dependence within each scheme, together with the difference between the two regularization schemes, provides an estimate of the theoretical uncertainty associated with the regularization procedure.

\subsection{In-medium mass at finite momentum}

\begin{figure}[t!]
    \centering
    \includegraphics[width=0.43\textwidth]{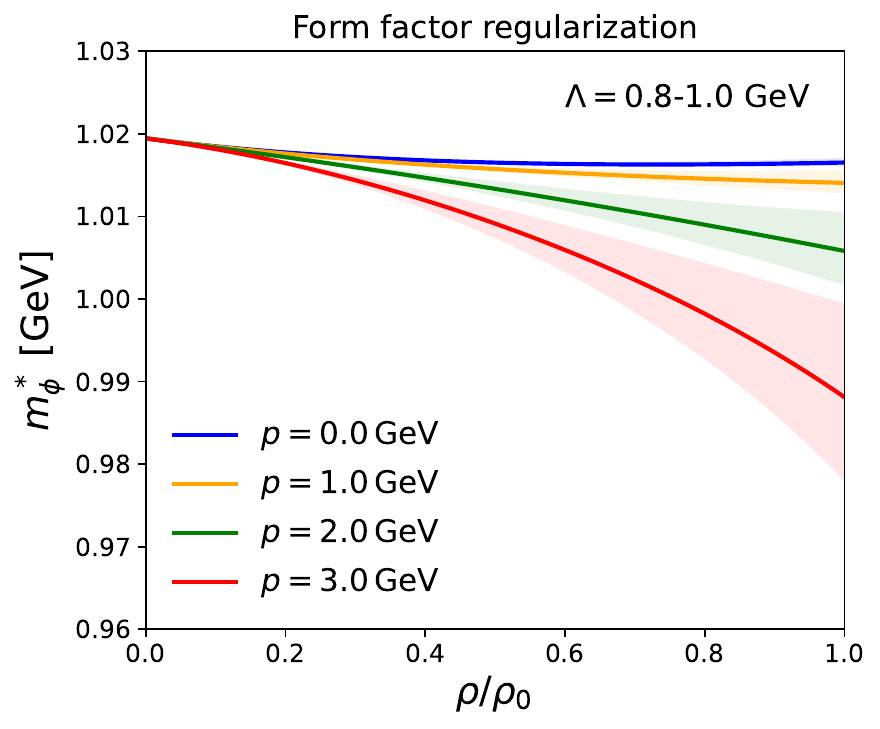}
    \includegraphics[width=0.43\textwidth]{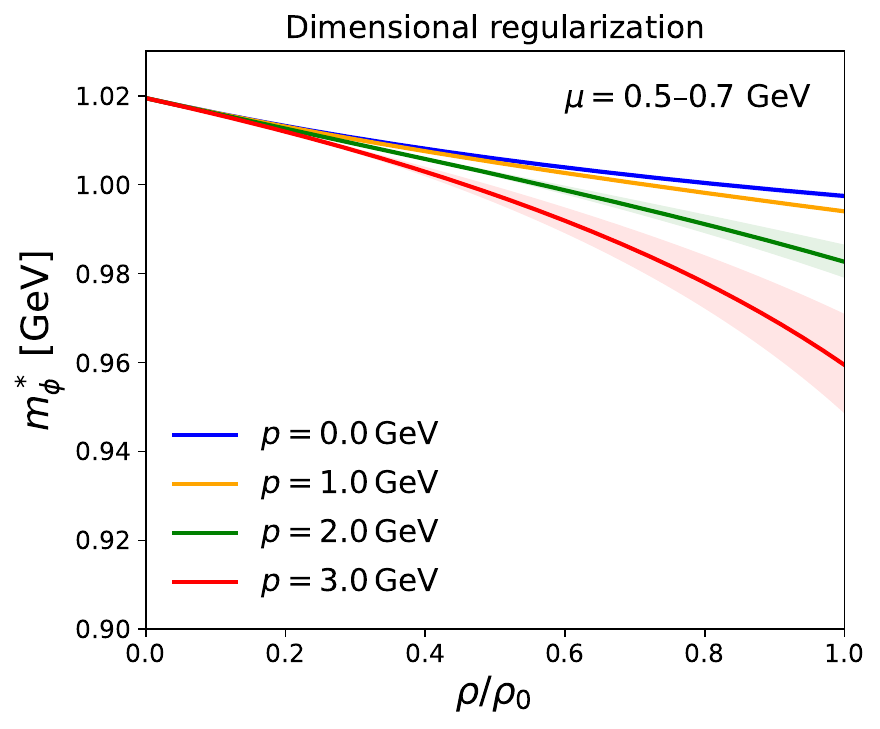}
    \caption{
    In-medium longitudinal mass of the $\phi$ meson as a function of nuclear density at momenta $p = 0, 1, 2,$ and $3~\mathrm{GeV}$,  computed using two different regularization scheme. The transverse mass remains identical to that of the $\phi$ meson at rest.
    }
    \label{fig:mass-lab}
\end{figure}

We now investigate the in-medium masses of the two polarization modes for a propagating $\phi$ meson in nuclear matter.
Figure~\ref{fig:mass-lab} shows the density dependence of the longitudinal-mode mass at fixed momenta $p = 0, 1, 2,$ and $3~\mathrm{GeV}$ by considering the loop and contact terms. Results obtained with two different regularization schemes display qualitatively similar behavior, although the cutoff dependence becomes more pronounced at finite momentum, as shown by the band, indicating enhanced sensitivity to ultraviolet contributions.

\begin{figure}[t!]
    \centering 
    \includegraphics[width=0.43\textwidth]{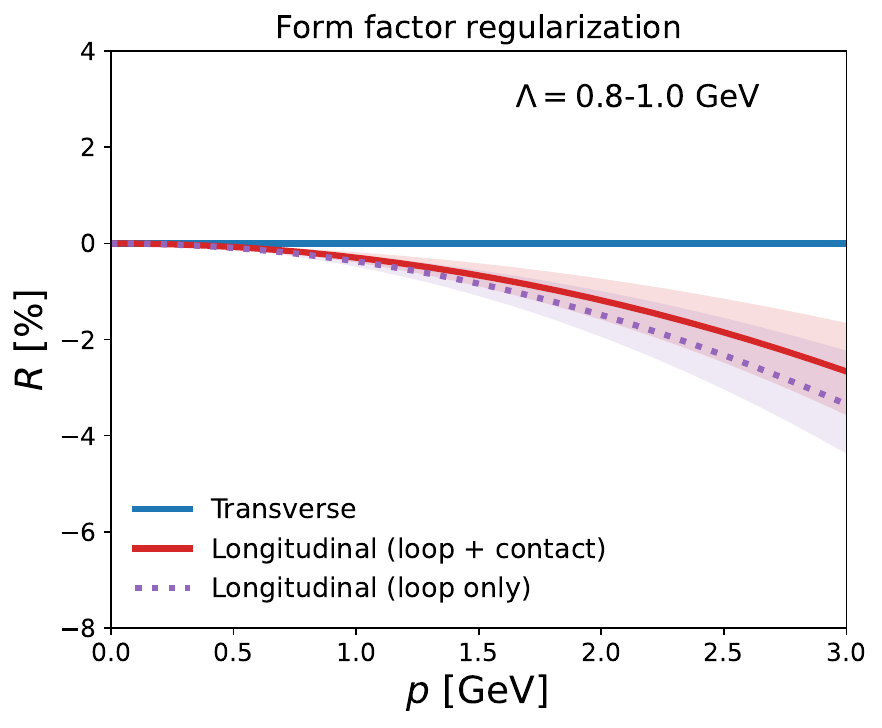}
    \includegraphics[width=0.43\textwidth]{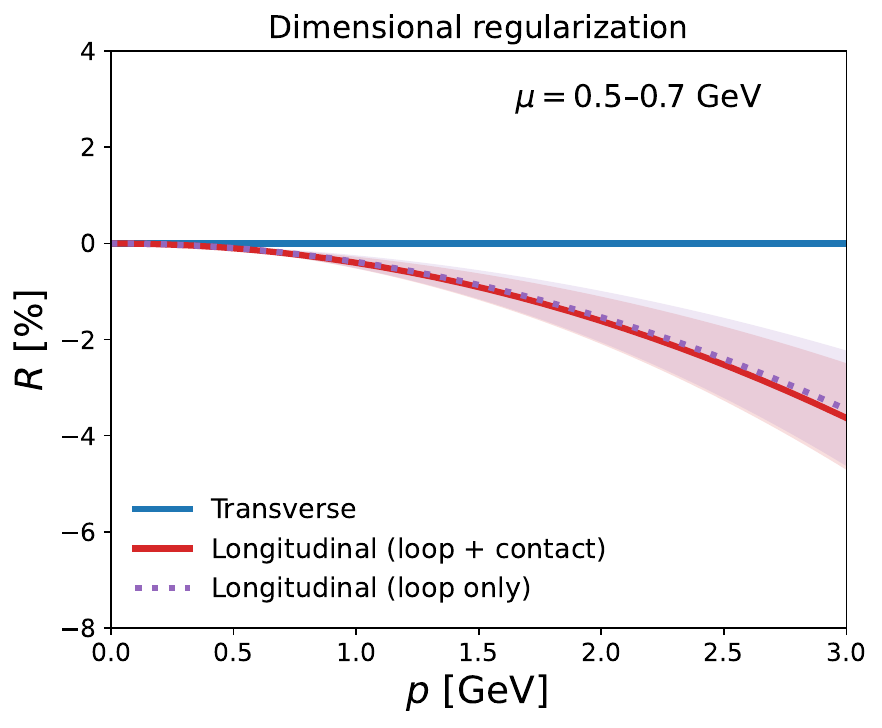}
    \caption{ 
    Momentum dependence $p$ of the in-medium $\phi$ meson mass ratio $R \equiv \Delta m_\phi^*(p)/m_\phi^*(0)$ (in percentage), where $\Delta m_\phi^*(p) = m_\phi^*(p) - m_\phi^*(0)$, at normal nuclear density for the longitudinal and transverse modes. The solid (dotted) lines correspond to calculations with (without) the contact term.
    }
    \label{fig:mass-momentum}
\end{figure}

Figure~\ref{fig:mass-momentum} shows how the in-medium mass shift evolves with momentum at normal nuclear density.
A clear difference between the two modes is observed: while the longitudinal mass decreases quadratically with increasing momentum $\bm{p}$, the transverse mass remains constant, showing no noticeable momentum dependence relative to its value at rest.
In the present approach, the transverse mode does not couple to the vector mean field at leading order [see Eq.~\eqref{eq:opt-t}]. The longitudinal mode, however, couples through derivative-type interactions [Eq.~\eqref{eq:opt-l}], which naturally generate a momentum-dependent contribution to the self-energy. 
As the $\phi$ meson momentum increases, this coupling strengthens the medium effect, leading to an enhanced separation between longitudinal and transverse modes. In vacuum, where the vector mean field vanishes, both polarizations remain degenerate for all momenta. 
We also find that the contact term leads to a slightly modified momentum dependence in the longitudinal mass ratio, while preserving the quadratic behavior.

For comparison, the behavior of the longitudinal polarization with increasing momentum is consistent with the QCDSR prediction~\cite{Kim:2019ybi}. However, the unchanged transverse mass differs from the QCDSR prediction of an approximately quadratic increase, although it carries larger uncertainties than the longitudinal mode. This mode separation is attributed to Lorentz-noninvariant condensates in nuclear matter entering the operator product expansion. Specifically, as shown in Fig.~3 of Ref.~\cite{Kim:2019ybi}, the longitudinal mode is dominated by a decreasing contribution, whereas the transverse mode receives competing contributions from quark and gluon condensates.

A direct correspondence between the two approaches is, however, not straightforward due to their different degrees of freedom and underlying approximations. While the QCDSR analysis is based on quark and gluon condensates, the present $K\bar{K}$-loop calculation provides an effective hadronic description of the dominant low-energy dynamics. Further studies, including additional interaction effects such as $\phi N$ resonance interactions~\cite{Peters:1997va} and coupled-channel contributions, are of great importance for understanding the origin of the differences between the two approaches and may in principle modify the results.

\subsection{In-medium width at finite momentum}

\begin{figure}[b!]
    \centering
    \includegraphics[width=0.43\textwidth]{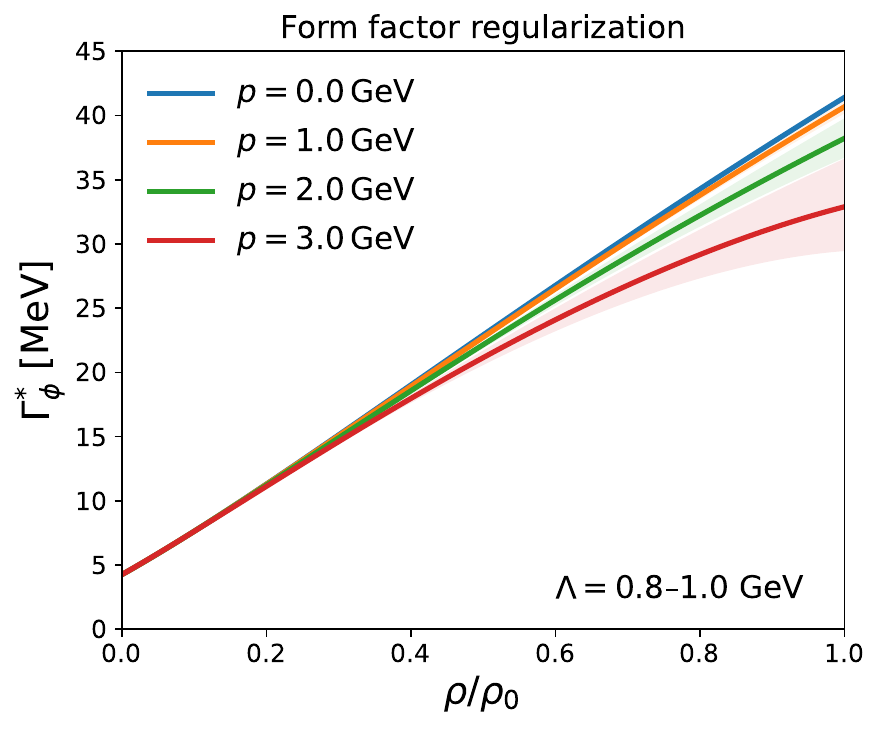}
    \includegraphics[width=0.43\textwidth]{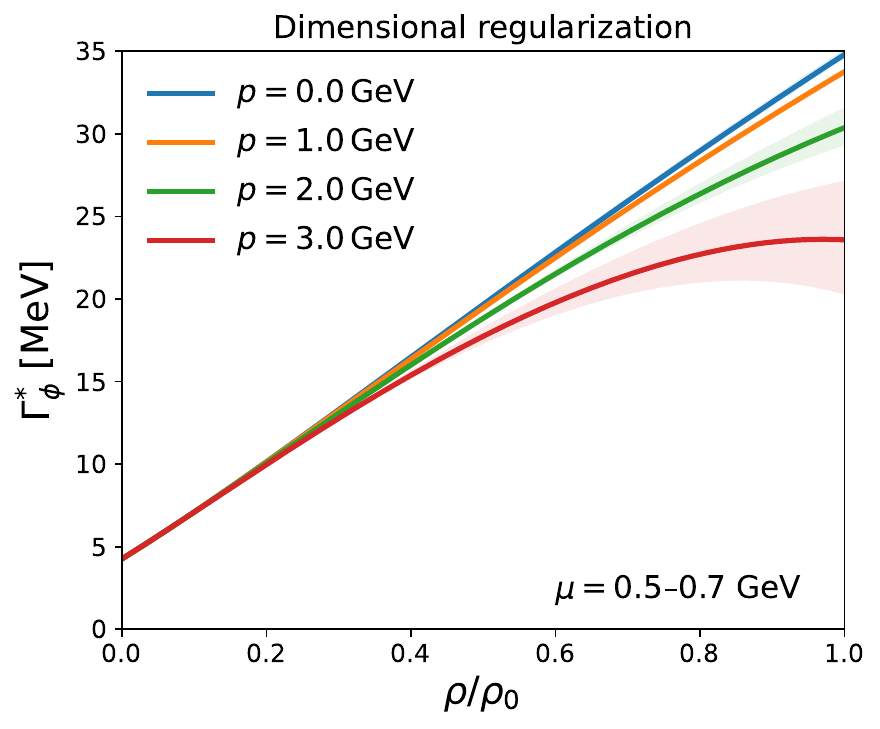}
\caption{
Density dependence of the intrinsic in-medium decay width $\Gamma_\phi^{*}$ of the longitudinal $\phi$ meson calculated using two different regularization schemes. The intrinsic width reflects the medium-induced broadening from the $\phi$-meson self-energy and is relevant for experimental observables. The intrinsic transverse width remains identical to that at rest.
}
    \label{fig:width-lab0}
\end{figure}

We next discuss the behavior of the in-medium intrinsic decay width, which is relevant for experimental observables.
Figure~\ref{fig:width-lab0} shows the density dependence of the longitudinal intrinsic decay width for several nonzero values of the $\phi$ meson momentum. The results are presented with both the loop and contact terms included.
Both schemes yield qualitatively consistent results, indicating that the predicted in-medium broadening remains unchanged against the choice of ultraviolet regularization, although quantitatively the results depend on the details of the cutoff and scale parameters. 

\begin{figure}[b]
    \centering
    \includegraphics[width=0.43\textwidth]{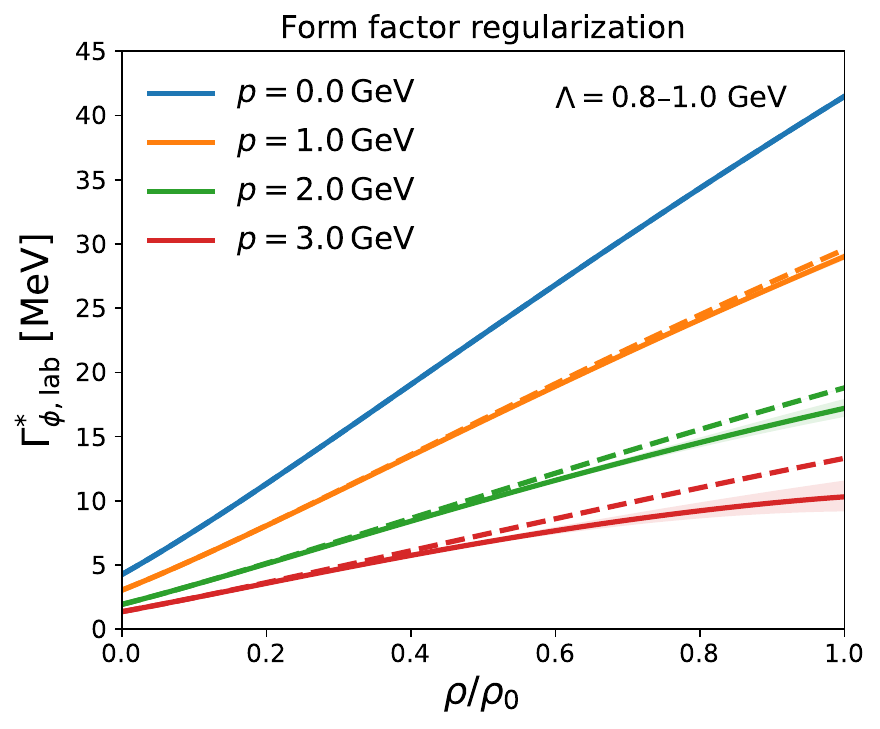}
    \includegraphics[width=0.43\textwidth]{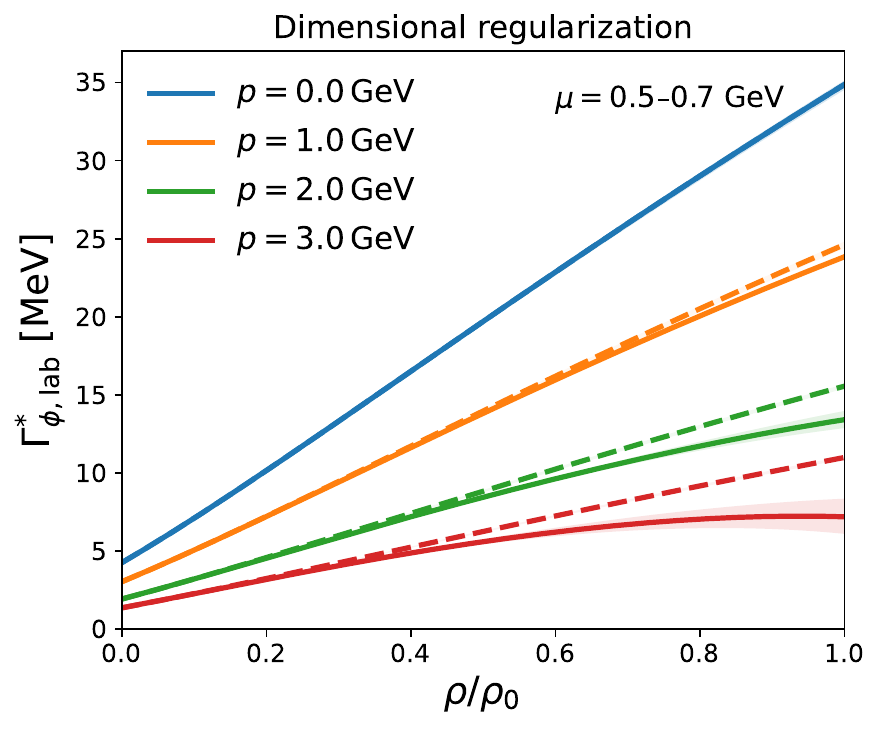}
    \caption{ Density dependence of the in-medium decay width $\Gamma_{\phi,\mathrm{lab}}^{*}$ of the longitudinal $\phi$ meson calculated using two different regularization schemes. The solid and dashed lines represent the longitudinal and transverse modes, respectively. This width includes Lorentz time dilation and corresponds to the decay rate in the laboratory frame. }
    \label{fig:width-lab-gamma}
\end{figure}

For $|\bm{p}|=0$, the width for both polarization modes increases significantly with increasing density. For finite momenta, the longitudinal width still increases with density, but its growth becomes progressively weaker as the $\phi$-meson momentum increases.
This behavior is primarily driven by the momentum-dependent modification of the longitudinal mass, which reduces the available decay phase space at higher momenta. 
In contrast, the explicit $\bm{p}^2$-dependent corrections in Eq.~\eqref{eq:width-p2} contribute only minor effects in this case.
The transverse width, on the other hand, remains essentially identical to its value at rest, consistent with the negligible momentum dependence of the transverse mass.

Figure~\ref{fig:width-lab-gamma} presents the corresponding results in the laboratory frame, including the Lorentz $\gamma$ factor, which governs the decay probability and survival of a moving $\phi$ meson in the nuclear medium. In vacuum, both the transverse and longitudinal decay widths decrease with increasing momentum due to Lorentz time dilation. Consequently, even in the absence of medium effects, the decay widths at finite momentum are reduced compared to their rest-frame values. In the nuclear medium, the competition between intrinsic in-medium broadening and Lorentz time dilation leads to a nontrivial momentum dependence of the decay widths. As a result, at finite momenta, the in-medium transverse decay width also differs from its rest-frame value, despite the absence of intrinsic momentum dependence in the transverse self-energy. The separation between longitudinal and transverse widths becomes increasingly pronounced at higher densities and momenta, reflecting the polarization-dependent nature of the medium modifications.

\subsection{Unpolarized spectral functions}

\begin{figure*}[t]
    \centering
    \includegraphics[width=0.95\textwidth]{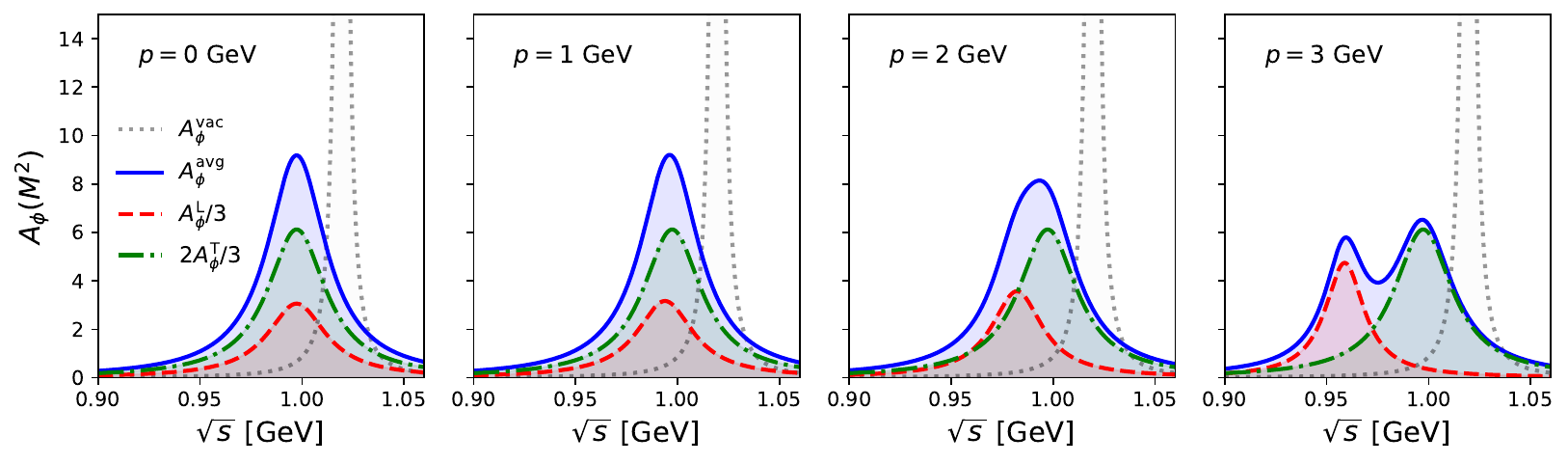}
    \caption{Unpolarized spectral functions of the $\phi$ meson at normal nuclear density for momenta $p = 0, 1, 2,$ and $3~\mathrm{GeV}$, calculated using dimensional regularization. The gray dotted lines represent the vacuum results, while the blue line corresponds to the in-medium spectral function. The individual longitudinal and transverse contributions, $A_\phi^{L}/3$ and $2A_\phi^{T}/3$, are also displayed.}
    \label{fig:spectum}
\end{figure*}

Finally, we examine the unpolarized spectral functions of the $\phi$ meson at normal nuclear density $\rho_0$, shown in Fig.~\ref{fig:spectum}. The gray dotted lines represent the vacuum spectral functions, while the blue lines show the in-medium results.
For demonstration, we present the results obtained using dimensional regularization $(\mu=0.6)$.

At $|\bm{p}|=0$, the longitudinal and transverse spectral functions are identical. The nuclear medium induces substantial broadening and a moderate downward mass shift, resulting in a significant redistribution of spectral strength toward lower energies.
For propagating $\phi$ mesons, $|\bm{p}| = 2$ and $3~\mathrm{GeV}$, the longitudinal and transverse spectral functions begin to separate due to polarization-dependent medium effects. However, because the two modes still strongly overlap, the resulting unpolarized spectral function exhibits a single broad peak. At $|\bm{p}| = 2~\mathrm{GeV}$, the mode separation appears only as a weak shoulder, while a distinct two-peak structure becomes visible at $|\bm{p}| = 3~\mathrm{GeV}$.

Compared to the QCDSR analysis~\cite{Kim:2019ybi}, an important difference emerges. In Ref.~\cite{Kim:2019ybi}, the transverse mass increases approximately quadratically with momentum, leading to a visible separation of the longitudinal and transverse peaks already at $|\bm p| \simeq 1.5$--$2.0~\mathrm{GeV}$. In contrast, the transverse mass remains nearly unchanged in the present calculation, and a distinct double-peak structure develops only at larger momenta, around $|\bm p| \simeq 2$--$3~\mathrm{GeV}$. This conclusion is consistent with the KEK-E325 results~\cite{KEK-PS-E325:2005wbm,KEK-PSE325:2025fms}, where the strongest medium effects were observed for slow $\phi$ mesons with $\beta\gamma = |\bm p|/m_\phi \lesssim 1.25$. In this momentum region, the longitudinal and transverse peaks remain strongly overlapping in the present calculation, implying that the polarization splitting is unlikely to be resolved through the unpolarized invariant-mass spectrum. Experimentally, the negative mass shift appears as an excess on the low-mass side of the dominant vacuum $\phi$ peak.

At higher momenta, corresponding to $\beta\gamma \gtrsim 1.75$ in the KEK-E325 kinematics, the separation between the longitudinal and transverse modes becomes more pronounced. However, the medium-induced modification becomes less visible, as reflected by the absence of a significant excess on the low-mass side of the $\phi$ spectrum~\cite{KEK-PS-E325:2005wbm,KEK-PSE325:2025fms}, because a larger fraction of $\phi$ mesons decay outside the nucleus. In the upcoming J-PARC experiment~\cite{Aoki:2023qgl}, momentum bins extending to $\beta\gamma \gtrsim 2.5$ are planned, where the two-peak structure may become more apparent. Nevertheless, since the probability of in-medium decay is further reduced at such momenta, the corresponding excess may remain small. Although the present results do not suggest a clear momentum window for observing a two-peak structure in the unpolarized spectrum, a more quantitative assessment can be achieved through dedicated transport calculations~\cite{Gubler:2024ovg} incorporating the momentum-dependent mass splitting.

\section{Conclusion and Outlook} 
\label{sec:conclusion}

In this work, we studied the in-medium properties of the $\phi$ meson in the quark--meson coupling (QMC) model~\cite{Tsushima:1997df}, incorporating self-consistent kaon mass modifications and $K\bar{K}$-loop contributions~\cite{Cobos-Martinez:2017vtr}. 
We explored the density and momentum dependence of the $\phi$ meson mass, width, and spectral functions, emphasizing polarization effects and assessing model dependence using covariant form factor and dimensional regularization schemes.

At normal nuclear density, the kaon effective mass decreases by about 13\%~\cite{Tsushima:1997df}, inducing a moderate downward mass shift of the $\phi$ meson by a few percent and a broadening of its decay width by an order of magnitude.
The qualitative results remain unchanged under variations in the regularization scheme, although quantitative differences are observed, particularly when the contact term is included. Nevertheless, after including the contact term, the predictions obtained using the form factor and dimensional regularization are consistent with those reported in Ref.~\cite{Cobos-Martinez:2017vtr} and Ref.~\cite{Ko:1992tp}, respectively.
At finite momentum, a pronounced polarization dependence emerges. 
While the transverse mode remains essentially unchanged with momentum, the longitudinal mode exhibits a quadratic decrease of the in-medium mass. The decay width also shows a nontrivial momentum and polarization dependence in nuclear matter. 

This polarization dependence of the mass shift can be understood from the tensor structure of the $\phi K\bar K$ interaction [Eq.~\eqref{eq:lagrangian1}] together with the polarization projectors $P_{\mu\nu}^{L,T}$. The transverse projector selects components perpendicular to the $\phi$-meson momentum, removing any explicit $|\bm p|$ dependence of the derivative vertex $(2q-p)^\mu$, whereas the longitudinal projector mixes temporal and longitudinal components, allowing momentum-dependent contributions to survive. Consequently, only the longitudinal mode acquires medium effects, while the transverse mode remains momentum independent.  
We further note that these results are obtained within the present hadronic $K\bar K$-loop model, and additional mechanisms such as $\phi N$ resonance interactions and coupled-channel dynamics may modify the momentum dependence. 

The predicted polarization dependence leads to a nontrivial momentum-dependent structure in the unpolarized spectral function. Although the polarization splitting is unlikely to be resolved directly through the unpolarized invariant-mass spectrum, polarization-sensitive observables, such as angular correlations in the $\phi\to K\bar K$ decay, may provide a direct probe of the longitudinal and transverse modes. Future measurements at J-PARC, particularly E88/Sa$\phi$re~\cite{Sako:2024oxb}, will be important for testing these predictions and discriminating between different theoretical descriptions of the in-medium $\phi$ meson. Similar polarization splitting has also been predicted for the $K_1^\pm$ meson~\cite{Park:2024vga,Yeo:2024iqz}, suggesting a broader manifestation of medium-induced Lorentz-symmetry breaking in hadronic systems.

\section*{Acknowledgment}

This work was supported by the RCNP Collaboration Research Network Program under Project No. COREnet 057. 
A.J.A. benefited from discussions at the Reimei Workshop on In-Medium Modification of Vector Mesons held at Yonsei University and acknowledges the support and hospitality.
A.J.A. was supported by the JAEA Postdoctoral Fellowship Program and partly by the PUTI Q1 Grant from the University of Indonesia under Contract No. PKS-206/UN2.RST/HKP.05.00/2025. 
A. J. A. is also supported by JSPS KAKENHI Grant Number 26K17163.
K.T. was supported by Conselho Nacional de Desenvolvimento Científico e  Tec-nológico (CNPq, Brazil), Processes No. 304199/2022-2, and FAPESP Process No. 2023/07313-6, and his work was also part of the projects, Instituto Nacional de Ciência e Tecnologia, Nuclear Physics and Applications (INCT-FNA), Brazil, Processes No. 464898/2014-5 and No 408419/2024-5. 
P.G. is supported by JSPS KAKENHI Grant Numbers JP22H00122 and JP25H00400.

\appendix  

\section{Form factor regularization}
\label{app:ff}

In this appendix, we provide the calculational details of the self-energy with the form factor regularization. Especially, the separate contributions of the loop and contact terms for both longitudinal and transverse polarizations. 
In addition, we also provide the explicit analytical expressions. 

For the loop contribution $\Pi^\lambda_{\rm loop}$ given in Eq.~\eqref{eq:se-ff}, the propagators in the self-energy can be rewritten as
\begin{align}
    \frac{1}{ D_\Lambda^2D_{\bar{K}}^*D_K^*} =&~ 6\int_0^1 \dd x \int_0^{1 - x} \dd y \nonumber\\
    &\times \frac{(1 - x - y)}{[D_\Lambda + x(D_{\bar{K}}^* - D_\Lambda) + y(D_K^* - D_\Lambda)]^4}\nonumber\\
    =&~ 6\int_0^1 \dd x \int_0^{1 - x} \dd y \frac{(1 - x - y)}{[(q-xp)^2 +C+i\epsilon]^4},
\end{align}
where 
\begin{align}
  C = x(1-x)m_\phi^{*2} - (x+y) m_{K}^{*2} - (1-x-y)\Lambda^2. 
\end{align}
To complete the square and ensure that the integration depends only on the $K\bar{K}$ loop momentum, we perform the shift $q^\mu \to q^\mu + xp^\mu.$
Using these ingredients, the $\phi$-meson self-energy becomes
\begin{align}\label{eq:pi_finite}
    \Pi^\lambda_{\rm loop}(m_\phi^{*2},\bm{p}^2) =&~ 12ig_\phi^2(\Lambda^2 -m_K^{*2})^2 \int_0^1 \dd x \int_0^{1 - x} \dd y\nonumber\\
    &~ \times (1 - x - y) \int \frac{d^4 q}{(2\pi)^4}
    \frac{ \mathcal{O}_{\rm loop}^{\lambda*}}{({q}^2 + C)^4}.
\end{align}
We note that the denominator contains no explicit $|\bm{p}|$ dependence; 
any momentum dependence arises solely from the numerator $\mathcal{O}_{\rm loop}^{\lambda*}$.

For the transverse mode, Eq.~\eqref{eq:pi_finite} reduces to
\begin{align}
    \Pi_\mathrm{loop}^T(m_\phi^{*2},\bm{p}^2) =&~ 24ig_\phi^2 (\Lambda^2 -m_K^{*2})^2\int_0^1 \dd x \int_0^{1 - x} \dd y~\nonumber\\
    & \times (1 - x - y) \left[\int \frac{d^4 q}{(2\pi)^4} \frac{ \bm{q}_\perp^{2}}{({q}^2 + C)^4}\right].
\end{align}
After performing the Wick rotation and carrying out the four-momentum integration in Euclidean space, 
the self-energy becomes
\begin{align}
   \Pi_\mathrm{loop}^T(m_\phi^{*2},\bm{p}^2) = N \int_0^1 \dd x \int_0^{1 - x} \dd y~\frac{(1 - x - y)}{C}, \label{eq:repi_T}
\end{align}
with $N={g_\phi^2(\Lambda^2 -m_K^{*2})^2 }/{4\pi^2}$.
For the longitudinal mode, the self-energy reads
\begin{align}
    &\Pi_\mathrm{loop}^L(m_\phi^{*2},\bm{p}^2) \nonumber\\
    &~= \frac{48ig_\phi^2(\Lambda^2 -m_K^{*2})^2}{m_\phi^{*2}} \int_0^1 \dd x \int_0^{1 - x} \dd y~ (1 - x - y) \nonumber\\
    &~~~ \times \left\{\int \frac{d^4 q}{(2\pi)^4} \frac{\left[ (q^0-V_\omega^K) |\bm{p}| - q^z E_\phi^* \right]^2}{({q}^2 + C)^4} \right\}.
\end{align}
Following the same procedure, we obtain
\begin{align}
    \Pi_\mathrm{loop}^L(m_\phi^{*2},\bm{p}^2) =&~ N \int_0^1 \dd x \int_0^{1 - x} \dd y~ (1 - x - y) \nonumber\\
    &\times \left( \frac{1}{C} - \frac{2 |\bm{p}|^2(V_\omega^K)^2}{m_\phi^{*2}}\frac{1}{C^2} \right). \label{eq:repi_L}
\end{align} 
Here, a quadratic dependence on the $\phi$-meson momentum appears, originating from the derivative coupling in the $\phi KK$ interaction.
We emphasize that the second term vanishes both in the vacuum $(V_\omega^K=0)$ and in the rest frame $(\bm{p}=0)$.
In these limits, the longitudinal and transverse self-energies coincide, ensuring the correct limiting behavior.

Next, we consider the contribution from the contact diagram, $\Pi^\lambda_{\rm con}$, which is given by
\begin{align}
    \Pi^\lambda_\mathrm{con}(m_\phi^{*2},\bm{p}^2)
    =&~ 4ig_\phi^2 (\Lambda^2 -m_K^{*2})^2 \int \frac{\dd^4 q}{(2\pi)^4} \frac{1 }{D_\Lambda^2D_K^*}.
\end{align}
This contribution is identical for both transverse and longitudinal polarizations.
The propagators can be rewritten as 
\begin{align}
    \frac{1 }{D_\Lambda^2 D_K^*} =&~ 2 \int_0^1 \dd x \,
   \frac{x}{\left[ x D_\Lambda + (1-x) D_K^* \right]^3}\\
        =&~2 \int_0^1 \dd x \,
   \frac{x}{\left( q^2 + E \right)^3},
\end{align}
with $E=-x\Lambda^2 - (1-x) m_K^{*2}$.
The self-energy is then given by
\begin{align}
        &\Pi^\lambda_\mathrm{con}(m_\phi^{*2},\bm{p}^2)\nonumber\\
        &~~= 8ig_\phi^2 (\Lambda^2 -m_K^{*2})^2  \int_0^1 \dd x\int \frac{\dd^4 q}{(2\pi)^4} \frac{x}{\left( q^2 + E\right)^3}.
\end{align}
After Feynman parameterization and Wick rotation, the contact contribution becomes
\begin{align}
        \Pi^\lambda_\mathrm{con}(m_\phi^{*2},\bm{p}^2) = 
        -\frac{g_\phi^2(\Lambda^2 -m_K^{*2})^2}{4\pi^{2}} \int \dd x \frac{x}{E}.
\end{align}

\subsection{Analytical expressions}
\label{app:anl-ff}

Here, we provide the analytic expressions for the form factor regularization.

\paragraph{Contact term}
The contact term in Eqs.~\eqref{eq:ffT} and \eqref{eq:ffL} is computed as
\begin{align}
        \Pi^\lambda_\mathrm{con}(m_\phi^{*2},\bm{p}^2) = -N \int \dd x\frac{x  } {-x\Lambda^2-(1-x)m_K^{*2}},
\end{align}
where the analytic result is given by
\begin{align}
     &\Pi^\lambda_\mathrm{con}(m_\phi^{*2},\bm{p}^2)\nonumber\\
     &~~=  \frac{g_\phi^2(\Lambda^2 -m_K^{*2})}{4\pi^{2}}\left[ 1 - \frac{m_K^{*2}}{\Lambda^2-m_K^{*2}} \ln\!\left(\frac{\Lambda^2}{m_K^{*2}}\right) \right].
\end{align}

\paragraph{$\bm{p}$-independent loop term} 
The self-energy of the loop term common to the longitudinal and transverse modes in Eqs.~\eqref{eq:ffT} and \eqref{eq:ffL}
for the $\bm{p}$-independent part is given by
\begin{align}
   \Pi^{T/L}_\mathrm{loop}(m_\phi^{*2},\bm{p}^2) = N \int_0^1 \dd x \int_0^{1 - x} \dd y~\frac{(1 - x - y)}{C}. 
\end{align}

In the region of $2m_K^* < m_\phi^* < \Lambda + m_K^*$, we obtain the real part of the self-energy analytically as
\begin{align}
   & \mathrm{Re}~\Pi^{T/L}_\mathrm{loop}(m_\phi^{*2},\bm{p}^2)\nonumber\\
   &~ =\frac{N }{12m_\phi^2\delta^2}\biggl\{ \Theta_1 + 2 \beta^3\ln\left(\frac{1+\beta}{1-\beta}\right) \nonumber\\
   &~~~ +  \gamma \Theta_2 \left[\tan^{-1}\left(\tfrac{\delta_+}{\gamma}\right) + \tan^{-1}\left(\tfrac{\delta_-}{ \gamma}\right)\right] + \Theta_3\ln\frac{\delta_K}{\delta_L}
   \biggr\},
\end{align}
where we define
\begin{align}
\delta_K=&~ {m_K^{*2}}/{m_\phi^{*2}},\qquad \delta=\delta_K-\delta_L, \\\
\delta_L=&~ {\Lambda^{2}}/{m_\phi^{*2}}, \qquad  \delta_\pm =~ 1\pm\delta,    
\end{align}
and
\begin{align}
    \beta =&~ \sqrt{1-4\delta_K},\\
    \gamma =&~ \sqrt{-1 + 2(\delta_K+\delta_L)-(\delta_K-\delta_L)^2}. 
\end{align}
To simplify further, we also define
\begin{align}
    \Theta_1 =&~ 2\delta (1-2\delta), \\
    \Theta_2 =&~ 2(2\delta^2 + 5\delta_K-\delta_L-1),\\
    \Theta_3 =&~ 2\delta^3 + 3\delta^2-6\delta_K+1.
\end{align}
In the same region of $2m_K^* < m_\phi^* < \Lambda + m_K^*$, the imaginary part is obtained as
\begin{align}
    \mathrm{Im}~\Pi^{L/T}_\mathrm{loop}(m_\phi^{*2},\bm{p}^2) =&~ - \frac{g_\phi^2}{24\pi} m_\phi^{*2} \beta^3.
\end{align}

\paragraph{$\bm{p}$-dependent loop term} 
Now, we compute the second integral in Eq.~\eqref{eq:ffL}, that is related to the $\bm{p}$-dependent part for the longitudinal mode, defined as $\Pi^{L2}_\mathrm{loop}$.
The self-energy is given by
\begin{align}
   \Pi^{L2}_\mathrm{loop}(m_\phi^{*2},\bm{p}^2) 
= -N_2 |\bm{p}|^2 \int_{0}^{1} \dd x \int_{0}^{\,1-x} \dd y \;
   \frac{(1 - x - y)}{C^2},
\end{align}
with $N_2 = {2N (V_\omega^K)^2}/{m_\phi^{*2}}$.

In the region of $2 m_K^* < m_{\phi}^*  < \Lambda + m_K^*$, the real part is analytically given by 
\begin{align}
 &\mathrm{Re}\,\Pi^{L2}_\mathrm{loop}(m_\phi^{*2},\bm{p}^2)\nonumber\\
 &~~= 
-\frac{N_2 |\bm{p}|^2}{ m_\phi^{*4}\delta^2}\Biggl\{
-\beta\ln \left( \frac{1+ \beta}{1-\beta} \right) - \frac{1}{2} \ln \frac{\delta_K}{\delta_L}   \nonumber \\
&~~~~~ - \frac{ \Theta_4}{ \gamma}  \Bigl[\tan^{-1}\left(\tfrac{\delta_+}{\gamma}\right)  + \tan^{-1}\left(\tfrac{\delta_-}{\gamma}\right) \Bigr] \Biggr\},
\end{align} 
with $\Theta_4= 1- 3\delta_K-\delta_L$. 
In the same region, the imaginary part is obtained as
\begin{equation} 
\mathrm{Im}\,  \Pi^{L2}_\mathrm{loop}(m_\phi^{*2},\bm{p}^2) = - \frac{g_\phi^2}{2\pi} \frac{\beta (V_\omega^K)^2}{{m_\phi^{*2}}}|\bm{p}|^2.
\end{equation}

\section{Dimensional regularization}
\label{app:dim-reg}

In this approach, the $\phi$-meson self-energy can be written as
\begin{align}
\Pi_\mathrm{total}^{\lambda}(m_\phi^{*2},\bm{p}^2) =&~ 2i g_\phi^2 \int \frac{\dd^4q}{(2\pi)^4} 
\left( \frac{\mathcal{O}_{\rm loop}^{\lambda*}}{D_K^* D_{\bar K}^*} + \frac{\mathcal{O}_{\rm con}^{\lambda*}}{D_K^*} \right),
\end{align}
without the need of a form factor.
For the loop contribution, we introduce Feynman parameters,
\begin{align}
\frac{1}{D_K^* D_{\bar K}^*} = \int_0^1 \dd x \frac{1}{\left[ x D_K^* + (1-x) D_{\bar K}^* \right]^2},
\end{align}
and shift the loop momentum $q^\mu \to q^\mu + (1-x)p^\mu$, which leads to
\begin{align}
\Pi_\mathrm{loop}^{\lambda}(m_\phi^{*2},\bm{p}^2) = 2i g_\phi^2 \int_0^1 \dd x \int \frac{\dd^4 q}{(2\pi)^4} 
\frac{\mathcal{O}_{\rm loop}^{\lambda*}}{(q^2 - \Delta^* + i\epsilon)^2}, 
\end{align}
where $\Delta^* = m_K^{*2} - x(1-x)m_\phi^{*2}$. 
For the contact term, the Feynman parameter is not needed.

\subsection{Transverse mode}
Including both the loop and contact contributions, the transverse self-energy can be written as
\begin{align}
    \Pi_\mathrm{total}^T(m_\phi^{*2},\bm{p}^2) 
    =&~ 4ig_\phi^2 \int \frac{\dd^d {q}}{(2\pi)^d}  \biggl( \int_0^1 \dd x 
       \frac{P_{\mu\nu}^T {q}^{\mu}{q}^{\nu}}{D_\Delta^{*2}} + \frac{1}{D_K^*} \biggr),
\end{align}
where, working in $d$ dimensions, we have used the identity $\bm{q}_\perp^{2} = P_{\mu\nu}^Tq^\mu q^\nu$.
The momentum integrals are evaluated using dimensional regularization in $d = 4 - \epsilon$ dimensions.
Using the standard integral relations
\begin{align}
\int \frac{\dd^d q}{(2\pi)^d} 
\frac{q^{\mu}q^\nu}{(q^2 - \Delta)^2}
=&~ \frac{-ig^{\mu\nu}}{2(4\pi)^{d/2}}\,
\Gamma\!\left(1 - \frac{d}{2}\right)\,
\Delta^{\frac{d}{2}-1},\\
\int \frac{\dd^d q}{(2\pi)^d} \frac{1}{q^2 - \Delta}
=&~ \frac{-i}{(4\pi)^{d/2}} \Gamma\left(1 - \frac{d}{2}\right) \Delta^{\frac{d}{2}-1},
\end{align}
the contact and loop contributions to the transverse self-energy become
\begin{align}
\Pi_\mathrm{con}^{T}(m_\phi^{*2},\bm{p}^2)
&= \frac{4g_\phi^2}{(4\pi)^{d/2}} \Gamma\left(1 - \frac{d}{2}\right) (m_K^{*2})^{\frac{d}{2}-1}. \\
\Pi_\mathrm{loop}^T(m_\phi^{*2},\bm{p}^2)
&= -\frac{4g_\phi^2\Gamma\!\left(1 - \frac{d}{2}\right)}{(4\pi)^{d/2}}\, \nonumber\\
&~~~~\times \left[\int_0^1 \dd x\, (\Delta^*)^{\frac{d}{2}-1} - (m_K^{*2})^{\frac{d}{2}-1}\right],
\end{align}
where we have used $g^{\mu\nu}P_{\mu\nu}^T=-2$.
Expanding around $d=4-\epsilon$, we use $(4\pi)^{-d/2} = (4\pi)^{-2}(1+\frac{\epsilon}{2} \ln 4\pi),$
$\Gamma\!\left(1 - \frac{d}{2}\right) = -\frac{2}{\epsilon} - 1 + \gamma_E$ and 
$\Delta^{\frac{d}{2}-1}= \Delta \left(1 - \frac{\epsilon}{2}\ln\Delta\right).$
Adding the contact term, the renormalized transverse self-energy reads
\begin{align} 
&\Pi_\mathrm{loop}^T(m_\phi^{*2},\bm{p}^2)\nonumber\\
&~ = -\frac{g_\phi^2}{4\pi^2}
    \int_0^1 \dd x\, \biggr(\Delta^*\ln\frac{\Delta^*}{\mu^2} - m_K^{*2} \ln \frac{m_K^{*2}}{\mu^2} - a(\mu)\frac{m_\phi^2}{6} \biggr), \label{eq:dim-t}
\end{align}
where the first two terms arise from the loop and contact contributions, respectively.
The real part of the transverse self-energy can be evaluated analytically as
\begin{align}
    &\text{Re}~\Pi_\mathrm{total}^T(m_\phi^{*2},\bm{p}^2) \nonumber\\
    &~~=
    -\frac{g_\phi^2 }{4\pi^2} \biggl\{ -\frac{m_\phi^{*2}}{6} \biggl[ \beta^3\ln\left( \frac{1+\beta}{1-\beta} \right) + \frac{8m_K^{*2}}{m_\phi^{*2}}  \nonumber\\
    &~~~~~ +  \ln\frac{m_K^{*2}}{\mu^2} -\frac{5}{3}\biggr]  - a(\mu)\frac{m_\phi^2}{6}   \biggr\}
\end{align}
with $\beta=\sqrt{1-4m_K^{*2}/m_\phi^{*2}}$.
Note that in Ref.~\cite{Ko:1992tp}, the constant proportional to $5/3$ was absorbed into $a(\mu)$. 
However, we retain it here for consistency with the numerical integration of the first term in Eq.~\eqref{eq:dim-t}. 

When we consider only the loop diagram without the contact term, the transverse self-energy is given by
\begin{align}
&\mathrm{Re}\,\Pi_{\mathrm{loop}}^{T}(m_{\phi}^{*2},\bm{p}^{\,2})\nonumber\\
&
=
-\frac{g_{\phi}^{2}}{4\pi^{2}}
\Biggl[
m_{\phi}^{*2}
\biggl(
-\frac{\beta^{3}}{6}
\ln\!\left(\frac{1+\beta}{1-\beta}\right)
+\frac{1-3\beta^{2}}{12}
\ln\!\left(\frac{m_{K}^{*2}}{\mu^{2}}\right)
\nonumber\\
&~~~~
+\frac{6\beta^{2}-1}{18}
\biggr)+a(\mu)\frac{(1-3\beta_{\rm vac}^{2})m_{\phi}^{2}}{12}\,\,
\Biggr],
\end{align}
where we note that the subtraction constant $a(\mu)$ is kept fixed at its vacuum value.

\subsection{Longitudinal mode}
 
For the longitudinal mode, the loop contribution is given by
\begin{align}
&\Pi_\mathrm{loop}^L(m_\phi^{*2},\bm{p}^2)\nonumber\\
&~~= \frac{8ig_\phi^2}{m_\phi^{*2}} \int_0^1 \dd x \int \frac{\dd^4 {q}}{(2\pi)^4} \frac{\left[ ({q}^0 - V_\omega^K) |\bm{p}| - {q}^z E_\phi^* \right]^2}{({q}^{2} - \Delta^*)^2}.
\end{align}
Since the cross term vanishes by symmetry, the expression reduces to
\begin{align}
{\mathcal{O}}^L_\mathrm{loop} = \frac{4}{m_\phi^{*2}} \biggl(({q}^z)^2 E_\phi^{*2} + ({q}^0)^2  |\bm{p}|^2 + |\bm{p}|^2 (V_\omega^K)^2\biggr).
\end{align}
After performing the Wick rotation and extending the integration to $d$ dimensions, we use
\begin{align}
    (q^z)^2 = \frac{1}{d} {q}^2_E, \hspace{1cm}
    (q^0)^2 = -\frac{1}{d} {q}^2_E,
\end{align}
where $q_E$ denotes the Euclidean four-momentum. This yields
\begin{align}
{\mathcal{O}}^L_\mathrm{loop} = \frac{m_\phi^{*2}}{d} {q}^2_E + |\bm{p}|^2 (V_\omega^K)^2.
\end{align}

The longitudinal loop contribution then becomes
\begin{align}
&\Pi_\mathrm{loop}^L(m_\phi^{*2},\bm{p}^2) \nonumber\\
&~~= -\frac{8g_\phi^2}{m_\phi^{*2}} \int_0^1 \dd x \int \frac{\dd^d {q}_E}{(2\pi)^d} \frac{\frac{m_\phi^{2}}{d}{q}^2_E + |\bm{p}|^2 (V_\omega^K)^2}{({q}^{2}_E + \Delta^*)^2}.
\end{align} 
Using the standard Euclidean integrals
\begin{align}
\int \frac{\dd^d q_E}{(2\pi)^d} \frac{1}{(q^2_E + \Delta)^2}
&= \frac{1}{(4\pi)^{d/2}} \Gamma\left(2 - \frac{d}{2}\right) \Delta^{\frac{d}{2}-2},\\
\int \frac{\dd^d q_E}{(2\pi)^d} \frac{q^2_E}{(q^2_E + \Delta)^2}
&= \frac{1}{(4\pi)^{d/2}} \frac{d}{2} \Gamma\left(1 - \frac{d}{2}\right) \Delta^{\frac{d}{2}-1}.
\end{align}
and combining the loop and contact contributions, we obtain
\begin{align}\label{eq:repi_L}
&\Pi_\mathrm{total}^L(m_\phi^{*2},\bm{p}^2)\nonumber\\
&~~= -\frac{g_\phi^2}{4\pi^2} \int_0^1 \dd x  \Biggl[ \Delta^* \ln\frac{\Delta^*}{\mu^2} - m_K^{*2} \ln \frac{m_K^{*2}}{\mu^2}  \nonumber\\
&~~~~~ - 2 (V_\omega^K)^2\frac{|\bm{p}|^2}{m_\phi^{*2}} \left( \ln\frac{\Delta^*}{\mu^2} + 1 \right) 
    - a(\mu)\frac{m_\phi^2}{6} \Biggr].
\end{align}
Note that the $\bm{p}$-dependent part also exhibits a $\mu$ dependence.
The real part can be evaluated analytically as
\begin{align}
    &\text{Re}~\Pi_\mathrm{total}^L(m_\phi^{*2},\bm{p}^2)\nonumber\\
    &~~= -\frac{g_\phi^2 }{4\pi^2} \biggl\{-\frac{m_\phi^{*2}}{6} \biggl[ \beta^3\ln\biggl( \frac{1+\beta}{1-\beta} \biggr) + \frac{8m_K^{*2}}{m_\phi^{*2}}  \nonumber\\
        &~~~~ +  \ln\frac{m_K^{*2}}{\mu^2} - \frac{5}{3} \biggr] -a(\mu)\frac{m_\phi^2}{6} \nonumber\\
        &~~~~ - 2(V_\omega^K)^2\frac{|\bm{p}|^2}{m_\phi^{*2}} \left(  \ln \frac{m_K^{*2}}{\mu^2} 
        - 1 + \beta\ln\frac{1+\beta}{1-\beta}\right)  \biggr\}. 
\end{align}
For the imaginary part, the results are identical to those obtained using the form-factor regularization, and we therefore do not discuss them further here.

Lastly, when we consider only the loop diagram without the contact term, the longitudinal self-energy reads
\begin{align}
&\mathrm{Re}\,\Pi_{\mathrm{loop}}^{L}(m_{\phi}^{*2},\bm{p}^{\,2})\nonumber\\
&=
-\frac{g_{\phi}^{2}}{4\pi^{2}}
\Biggl\{
m_{\phi}^{*2}
\biggl[
-\frac{\beta^{3}}{6}
\ln\!\left(\frac{1+\beta}{1-\beta}\right)
+\frac{1-3\beta^{2}}{12}
\ln\!\left(\frac{m_{K}^{*2}}{\mu^{2}}\right)
\nonumber\\
&~~~~
+\frac{6\beta^{2}-1}{18}
\biggr] + a(\mu)\frac{(1-3\beta_{\rm vac}^{2})m_{\phi}^{2}}{12}\,\, \nonumber\\
&~~~~ - 2(V_\omega^K)^2\frac{|\bm{p}|^2}{m_\phi^{*2}} \left(  \ln \frac{m_K^{*2}}{\mu^2} 
        - 1 + \beta\ln\frac{1+\beta}{1-\beta}\right)
\Biggr\}.
\end{align}

\bibliography{references}

\end{document}